\documentclass[epj]{svjour}
\usepackage{graphics}
\usepackage{fancyhdr}
\usepackage{amssymb}
\usepackage{epsfig}
\usepackage{amsmath}
\def\be{\begin{equation}}
\def\ee{\end{equation}}
\def\bea{\begin{eqnarray}}
\def\eea{\end{eqnarray}}
\def\nn{\nonumber}
\def\mz{\mathbb Z}

\newcommand{\lgp}[2]{{\rm {#1}_{#2}}}

\newcommand{\SU}[1]{\ensuremath{\mathrm{SU}(#1)}}
\newcommand{\BmL}{\ensuremath{B\!-\!L} }
\newcommand{\ov}{\overline}
\newcommand{\mc}{\mathcal}

\begin{document}
\title{SUSY GUT Model Building}
\author{Stuart Raby
\thanks{\emph{Dedicated to the memory of Julius Wess.} }%
}                     % Do not remove
\institute{The Ohio State University, 191 W. Woodruff Ave, Columbus,
OH 43210, USA \\ Preprint No. OHSTPY-HEP-T-08-003  }
\date{Received: date / Revised version: date}
\abstract{I discuss an evolution of SUSY GUT model building,
starting with the construction of 4d GUTs, to orbifold GUTs and
finally to orbifold GUTs within the heterotic string. This evolution
is an attempt to obtain realistic string models, perhaps relevant
for the LHC.   This review is in memory of the sudden loss of Julius
Wess, a leader in the field, who will be sorely missed.
\PACS{
      {PACS-key}{discribing text of that key}   \and
      {PACS-key}{discribing text of that key}
     } % end of PACS codes
} %end of abstract
\maketitle
\section{\label{sec:motivation} Motivation}

Before beginning this review on SUSY GUTs, it is probably worthwhile
spending a very brief moment motivating the topic. What are the
virtues of SUSY GUTs?  The following is a list of all the issues
that SUSY GUTs either addresses directly or provides a framework for
addressing.

\begin{enumerate}
\item $M_{Z} << M_{GUT}$ ``Natural"
\item Explains Charge Quantization and family structure
\item Predicts Gauge Coupling Unification$^*$
\item Predicts Yukawa Coupling Unification
\item + Family Symmetry $\Longrightarrow$ Hierarchy of Fermion Masses
\item Neutrino Masses via See - Saw scale $\sim 10^{-3} - 10^{-2} \ M_G$
\item LSP -- Dark Matter Candidate
\item Baryogenesis via Leptogenesis
\item SUSY Desert $\Longrightarrow$ LHC experiments probe \\ physics
O($M_{Planck}$) scale
\item SUSY GUTs are natural extension of the Standard Model
\end{enumerate}

In the following review we will discuss some of these issues in
great detail.   Let us start by defining our notation for the
Standard Model.

\section{\label{sec:SM} Standard Model}

Let us define the generators for the gauge group $SU(3) \otimes
SU(2) \otimes U(1)_Y$ as ${\cal T}_A, \ A = 1, \dots, 8$ for $SU(3)$
, $T_a, \ a = 1, 2, 3$ for $SU(2)$ and the hypercharge operator $Y$
for $U(1)_Y$.

The gauge interactions of the quarks and leptons of the Standard
Model are then completely defined in terms of their gauge quantum
numbers.    The quark and lepton fields for one family are given in
terms of the left-handed Weyl spinors - \be q = \left(
\begin{array}{c} u \\ d \end{array} \right), \  u^c, \  d^c, l=
\left(
\begin{array}{c} \nu \\ e \end{array} \right), \  e^c, \ \nu^c \ee with SM charges
given by

\bea {\cal T}_A \ q = \frac{1}{2} \lambda_A \ q, \ {\cal T}_A \  u^c
= - \frac{1}{2} \lambda_A^T \ u^c, \ {\cal T}_A \ d^c = -
\frac{1}{2} \lambda_A^T \ d^c  \nn \\
{\cal T}_A \ l = {\cal T}_A \ e^c = {\cal T}_A \ \nu^c = 0 \eea
(where $\lambda_A$ are the 3 $\times$ 3 Gell-Mann matrices).

\bea T_a \ q = \frac{1}{2} \tau_a \ q, \ T_a \ l = \frac{1}{2}
\tau_a \ l \nn \\  T_a \ u^c = T_a \ d^c = T_a \ e^c = T_a \ \nu^c =
0 \eea (where $\tau_a$ are the 2 $\times$ 2 Pauli matrices).

\bea Y \ q = \frac{1}{3} \ q, \ Y \ u^c = - \frac{4}{3} \ u^c, \ Y \
d^c = \frac{2}{3} \ d^c, \nn \\ \ Y \ l = - l, \ Y \ e^c = +2 e^c, \
Y \ \nu^c = 0 . \eea

With this notation, the gauge covariant derivative is given by \be
D_\mu = ( \partial_\mu + i g_s {\cal T}_A {\cal G}_{\mu A} + i g T_a
W_{\mu a} + i g^\prime \ \frac{Y}{2} B_\mu ) \ee  and the electric
charge operator is given by $Q = T_3 + \frac{Y}{2}$.

And the gauge-fermion Lagrangian is given by \be {\cal
L}_{gauge-fermion} = [ l^* \ i \bar \sigma_\mu D^\mu \ l + \cdots] +
[ - \frac{1}{2} Tr( {\cal G}_{\mu \nu}  {\cal G}^{\mu \nu} ) +
\cdots ] . \ee\footnote{$ {\cal G}_{\mu \nu} = \partial_\mu {\cal
G}_{\nu} - \partial_\nu {\cal G}_{\mu} + i [{\cal G}_{\mu},{\cal
G}_{\nu}]$ with ${\cal G}_{\mu} = T^f_A {\cal G}_{\mu A}$ and
$T^f_A$ is the generator in the fundamental representation.} In
order to make contact with phenomenology it is sometimes useful to
use Dirac four component notation.  For example, the Dirac 4
component electron field, in terms of the 2 component
Weyl spinors, is given by \be \Psi_e = \left( \begin{array}{c} e \\
i \sigma_2 \ (e^c)^* \end{array} \right) . \ee

In addition we must add the Higgs bosons.  We will introduce the
minimal set of Higgs doublets consistent with supersymmetry. \be H_u
= \left( \begin{array}{c} h^+ \\ h^0 \end{array} \right), H_d =
\left(
\begin{array}{c} \bar h^0 \\ \bar h^- \end{array} \right) \ee
satisfying \bea {\cal T}_A \ H_u = {\cal T}_A \ H_d = 0 \nn \\
T_a \ H_u = \frac{1}{2} \tau_a \ H_u, \ T_a \ H_d = \frac{1}{2}
\tau_a \ H_d \nn \\
Y \  H_u = + H_u, \ Y \  H_d = - H_d . \eea

\bea - {\cal L}_{Yukawa}  = & \lambda_e^{i j} \ l_i \ e^c_j \ H_d +
\lambda_d^{i j} \ q_i \ d^c_j \ H_d + &
\\ & \lambda_u^{i j} \ q_i \
u^c_j \ H_u +  \lambda_\nu^{i j} \ l_i \ \nu^c_j \ H_u - \frac{1}{2}
M_{i j} \ \nu^c_i \ \nu^c_j . & \nn \eea

The generalization to the Minimal Supersymmetric Standard Model is
then quite simple.   One defines the left-handed chiral superfields.
For example, the electron left-handed Weyl field, $e$, is contained
in the left-handed chiral superfield, $E$, with \be E(y, \theta) =
\tilde e(y) + \sqrt{2} (\theta \ e(y)) + (\theta \ \theta) F_e(y)
\ee where the product $(\theta \ e(y)) \equiv \theta^\alpha \
e(y)_\alpha$ and \\ $y^\mu = x^\mu - i \theta \ \sigma^\mu
\theta^*$.

Then the supersymmetric Lagrangian includes the gauge-matter terms
\bea & {\cal L}_{gauge-matter}  = & \\ & [ \int d^4\theta
L_i^\dagger \ \exp(- 2 \ [g_s V_g + g V_W + g^\prime \frac{Y}{2}
V_B]) \ L_i + \cdots]  & \nn \\ & +  [ \frac{1}{8 g_s^2} \int d^2
\theta Tr({{\cal W}_g}^\alpha \ {{\cal W}_g}_\alpha + h.c.) + \cdots
] & \nn \eea where $V_g = V_A \ {\cal T}_A, \ V_W = V_a \ T_a$ and
the flavor index $i = 1, 2, 3$.

${\cal L}_{Yukawa} = \int d^2 \theta W$ with the superpotential
given by  \bea & W = & \\ & \lambda_e^{i j} \ L_i \ E^c_j \ H_d +
\lambda_d^{i j} \ Q_i \ D^c_j \ H_d  + \lambda_u^{i j} \ Q_i \ U^c_j
\ H_u & \nn \\ & + \lambda_\nu^{i j} \ L_i \ N^c_j \ H_u -
\frac{1}{2} M_{i j} \ N^c_i \ N^c_j - \mu \ H_u \ H_d .& \nn \eea

\section{\label{sec:GUTs} Two roads to Grand Unification}

One can first unify quarks and leptons into two irreducible
representations of the group  $SU(4)_C \otimes SU(2)_L \otimes
SU(2)_R$, i.e.  the so-called Pati-Salam group \cite{ps} where
lepton number is the fourth color.

Then the PS fields \be {\cal Q} = ( q \; l ),  \;  {\cal Q}^c
= ( q^c \; l^c) \ee where \be q^c = \left( \begin{array}{c} u^c \\
d^c
\end{array} \right), \; l^c = \left( \begin{array}{c} \nu^c \\ e^c
\end{array} \right) \ee transform as $(4, 2, 1) \oplus (\bar 4,
1, \bar 2)$ under PS.  One can check that baryon number minus lepton
number acting on a 4 of $SU(4)$ is given by \be B - L =  \left( \begin{array}{cccc} \frac{1}{3} & & & \\
& \frac{1}{3} & & \\ & &  \frac{1}{3} & \\ & & & -1 \end{array}
\right) \ee  and similarly electric charge is given by \be Q = T_{3
L} + T_{3 R} + \frac{1}{2} (B - L) . \ee
Note, charge is quantized since it is embedded in a non-abelian
gauge group.  One family is contained in two irreducible
representations.  Finally, if we require parity ( $L \leftrightarrow
R$ ) then there are two independent gauge couplings.

What about the Higgs?   The two Higgs doublets $H_u, \ H_d$ are
combined into one irreducible PS Higgs multiplet \be {\cal H} = (H_d
\ H_u ) \ee  transforming as a   $(1, 2, \bar 2)$ under PS.   Thus
for one family, there is a unique renormalizable Yukawa coupling
given by \be \lambda \ {\cal Q}^c \ {\cal H} \ {\cal Q}  \ee  giving
the GUT relation \be \lambda_t = \lambda_b = \lambda_\tau =
\lambda_{\nu_\tau} \equiv \lambda . \ee

Now Pati-Salam is {\em not} a grand unified gauge group.  However,
since $SU(4) \approx SO(6)$ and $SU(2) \otimes SU(2) \approx SO(4)$
(where $\approx$ signifies a homomorphism), it is easy to see that
PS $\approx SO(6) \otimes SO(4) \subset SO(10)$ \cite{SO(10)}. In
fact one family of quarks and leptons is contained in the spinor
representation of $SO(10)$, i.e.  \bea SO(10) \rightarrow SU(4)_C
\otimes SU(2)_L \otimes SU(2)_R \nn \\ 16 \rightarrow  (4, 2, 1)
\oplus (\bar 4, 1, \bar 2) . \eea  Hence by going to $SO(10)$ we
have obtained quark-lepton unification ( one family contained in one
spinor representation ) and gauge coupling unification (one gauge
group) (see Table \ref{tab:spinor}).

But I should mention that there are several possible breaking
patterns for $SO(10)$. \bea SO(10) & \rightarrow
SU(4)_C \otimes SU(2)_L \otimes SU(2)_R  \\
& \rightarrow SU(5) \otimes U(1)_X \nn \\
& \rightarrow SU(5)^\prime \otimes U(1)_{X^\prime} \nn \\
& \rightarrow SU(3)_C \otimes U(1)_{(B-L)} \otimes SU(2)_L \otimes
SU(2)_R . \nn \eea
 In order to preserve a
prediction for gauge couplings we would require the breaking pattern
\be SO(10) \rightarrow SM \ee or \be SO(10) \rightarrow SU(5)
\rightarrow SM . \ee

\begin{table}
$$
\begin{array}{|l|c|c|c|}
\hline \multicolumn{4}{|l|}{\rm Grand \;\; Unification \; - \; SO(10)}\\
 \hline
 {\rm State} & { \rm Y}   & {\rm Color} & {\rm Weak} \\
 & =  \frac{2}{3} \Sigma ({\rm C}) - \Sigma ({\rm W}) &
 {\rm C \; spins} & {\rm W \; spins} \\
\hline\hline
{\bf \bar \nu}  & { 0} &  + \, + \, +  &     + \, +  \\
\hline\hline
{\bf \bar e}  & { 2} &  + \, + \, + & - \, -  \\
\hline
{\bf u_r} &  &  - \, + \, + &   + \, - \\
{\bf d_r} &  &  - \, + \, + &   - \, + \\
{\bf u_b} & { \frac{1}{3}} &  + \, - \, + &   + \, -  \\
{\bf d_b} &  &  + \, - \, + &  - \, +  \\
{\bf u_y} &  & + \, + \, - &    + \, - \\
{\bf d_y} &  & + \, + \, - &   - \, + \\
\hline
{\bf \bar u_r} &  & + \, - \, - & + \, + \\
{\bf \bar u_b} & { -\frac{4}{3}} & - \, + \, - & + \, + \\
{\bf \bar u_y} &  & - \, - \, + & + \, + \\
\hline\hline
{\bf \bar d_r} &  & + \, - \, - & - \, - \\
{\bf \bar d_b} & { \frac{2}{3}} & - \, + \, - & - \, - \\
{\bf \bar d_y} &  & - \, - \, + & - \, - \\
\hline
{\bf \nu}  & { -1} &  - \, - \, - & + \, - \\
{\bf e}  &  & - \, - \, - & - \, + \\
\hline
\end{array}
$$
\caption{Spinor representation of $SO(10)$ where this table
explicitly represents the Cartan-Weyl weights for the states of one
family of quarks and leptons.  The double lines separate irreducible
representations of $SU(5)$. \label{tab:spinor}}
\end{table}

It will be convenient at times to work with the Georgi-Glashow GUT
group $SU(5)$ \cite{gg}.  We have $16 \rightarrow 10 \oplus \bar 5
\oplus 1$. Let's identify the quarks and leptons of one family
directly. We define the group $SU(5)$ by \be SU(5) = \{U | U = 5
\times 5 \; {\rm complex \; matrix}; U^\dagger U = 1 ; \det U = 1 \}
\ee and the fundamental representation $5^\alpha,  \ \alpha = 1,
\dots, 5$ transforms as \be {5^\prime}^\alpha = {U^\alpha}_\beta \
5^\beta . \ee We represent the unitary matrix $U$ by  \be U = \exp(i
T_A \ \omega_A) \ee where $Tr(T_A) = 0,  \ T_A^\dagger = T_A$, $A =
1, \dots, 24$ and \\ $[T_A, \ T_B] = i f_{ABC} \ T_C$ with $f_{ABC}$
the structure constants of $SU(5)$. Under an infinitesimal
transformation, we have \be \delta_A 5^\alpha = i
{(T_A)^\alpha}_\beta \ 5^\beta \ \omega_A . \ee

Let us now identify the $SU(3) \otimes SU(2) \otimes U(1)_Y$
subgroup of $SU(5)$.   The $SU(3)$ subgroup is given by the
generators \be T_A = \left( \begin{array}{c|c} \frac{1}{2} \lambda_A
& 0 \\ \hline 0 & 0 \end{array} \right),  \;\; A = 1, \dots, 8 . \ee
And the $SU(2)$ subgroup is given by \be  T_A =  \left(
\begin{array}{c|c} 0 & 0 \\ \hline 0 & \frac{1}{2} \tau_{(A- 20)}
\end{array} \right),  \;\; A = 21, 22, 23.  \ee

The generators in $SU(5)/ SU(3)\otimes SU(2) \otimes U(1)_Y$ are
given by \be T_A, \ A = 9, \dots, 20 . \ee  These are 12 generators
of the form  \be \left( \begin{array}{c|c} 0 & \begin{array}{cc} 1 &
0 \\ 0 & 0 \\ 0 & 0 \end{array} \\ \hline \begin{array}{ccc} 1 & 0 &
0 \\ 0 & 0 & 0 \end{array} & 0 \end{array} \right), \;\; \left(
\begin{array}{c|c} 0 & \begin{array}{cc} - i & 0 \\ 0 & 0 \\ 0 & 0
\end{array} \\ \hline \begin{array}{ccc} i & 0 & 0 \\ 0 & 0 & 0
\end{array} & 0 \end{array} \right) . \ee

Let us now identify the hypercharge $Y$.  The only remaining
generator of $SU(5)$ commuting with the generators of $SU(3)$ and
$SU(2)$ is given by \be T_{24} = \sqrt{\frac{3}{5}} \left(
\begin{array}{c|c} \begin{array}{ccc} -1/3 & 0 & 0 \\ 0 & -1/3 & 0
\\ 0 & 0 & -1/3 \end{array} & 0 \\ \hline 0 & \begin{array}{cc} 1/2
& 0 \\ 0 & 1/2 \end{array} \end{array} \right) \equiv
\sqrt{\frac{3}{5}}  \ \frac{Y}{2} . \ee  The overall normalization
is chosen so that all the $SU(5)$ generators satisfy  $Tr(T_A \ T_B)
= \frac{1}{2} \ \delta_{A B}. $

With these identifications, we see that the quantum numbers of a \be
5^\alpha = \left( \begin{array}{c} d^a \\ l^i \end{array} \right),
\; a = 1, 2, 3; i = 4, 5 \ee where $d^a$ transforms as $(3, 1,
-2/3)$ and $l^i$ transforms as $(1, 2, +1)$ under the SM. Of course
these are not correct quantum numbers for any of the quarks and
leptons, but the charge conjugate states are just right.  We have
\be \bar 5_\alpha = \left( \begin{array}{c} d^c_a \\ l^c_i
\end{array} \right), \; a = 1, 2, 3; i = 4, 5 \ee  with
transformation properties \be d^c = (\bar 3, 1, 2/3),  \;\; l^c =
(1, 2, -1) \ee and \be l^c = \left( \begin{array}{c} -e \\ \nu
\end{array} \right) . \ee

Once we have identified the states of the $\bar 5$, we have no more
freedom for the $10$.  The $10$ transforms as an anti-symmetric
tensor product of two $5$s, i.e.  \be 10^{\alpha \beta} = -
10^{\beta \alpha} \propto 5_1^\alpha \ 5_2^\beta - 5_2^\alpha \
5_1^\beta . \ee We find \bea  10^{a b} \equiv \epsilon^{a b c}
(u^c)_c = (\bar 3, 1, -4/3) \nn \\ 10^{a i} \equiv q^{a i} = (3, 2,
1/3) \nn \\ 10^{i j} \equiv \epsilon^{i j} e^c = (1, 1, +2) . \eea

To summarize we find \be \bar 5_\alpha = \left( \begin{array}{c}
d^c_1 \\ d^c_2 \\d^c_3 \\ - e \\ \nu \end{array} \right), \;\;
10^{\alpha \beta} = \frac{1}{\sqrt{2}} \ \left( \begin{array}{c|c}
\begin{array}{ccc} 0 & u^c_3 & - u^c_2 \\ - u^c_3 & 0 &  u^c_1
\\u^c_2 & - u^c_1 & 0
\end{array} & \begin{array}{cc} u^1 & d^1 \\u^2 & d^2 \\u^3 & d^3
\end{array} \\ \hline \begin{array}{ccc} - u^1 & - u^2 & - u^3 \\- d^1 & - d^2 & -
d^3 \end{array} & \begin{array}{cc} 0 & e^c \\ -e^c & 0 \end{array}
\end{array} \right) . \ee

Now that we have identified the states of one family in $SU(5)$, let
us exhibit the fermion Lagrangian (with gauge interactions).  We
have \be {\cal L}_{fermion} = \bar 5_\alpha^\dagger i (\bar
\sigma_\mu \ {D^\mu )_\alpha}^\beta \ \bar 5_\beta  + {10^{\alpha
\beta}}^\dagger i (\bar \sigma_\mu \ D^\mu )^{\alpha \beta}_{\gamma
\delta} \ 10^{\gamma \delta} \ee where \be D^\mu = \partial^\mu + i
g_G T_A \ A^\mu_A  \ee  and $T_A$ is in the $\bar 5$ or $10$
representation.   We see that since there is only one gauge coupling
constant at the GUT scale we have  \be g_3 = g_2 = g_1 \equiv g_G
\ee where, after weak scale threshold corrections are included, we
have \be g_3 \rightarrow g_s, \; g_2 \rightarrow  g, \; g_1
\rightarrow \sqrt{\frac{5}{3}}  \ g^\prime . \ee  At the GUT scale
we have the relation \be \sin^2\theta_W = \frac{(g^\prime)^2}{g^2 +
(g^\prime)^2} = 3/8 . \ee

But these are tree level relations which do not take into account
threshold corrections at either the GUT or the weak scales nor
renormalization group [RG] running from the GUT scale to the weak
scale.  Consider first RG running.   The one loop RG equations are
given by \be \frac{d \alpha_i}{d t} = - \frac{b_i}{2 \pi} \alpha_i^2
\ee where $\alpha_i = \frac{g_i^2}{4 \pi},  \; i = 1, 2, 3$ and \be
\label{eq:RGgeneral} b_i = \frac{11}{3} C_2(G_i) - \frac{2}{3} T_R \
N_F - \frac{1}{3} T_R \ N_S . \ee  Note, $t = - \ln
(\frac{M_G}{\mu})$,  $\sum_A (T_A^2) = C_2(G_i) \mathbb{I}$ with
$T_A$ in the adjoint representation defines the quadratic Casimir
for the group $G_i$ with $C_2(SU(N)) = N$ and $C_2(U(1)) = 0$.
$Tr(T_A T_B) = T_R \ \delta_{A B}$ for $T_A$ in the representation
$R$ (for $U(1)_Y$, $T_R \equiv \frac{3}{5} \ Tr( \frac{Y^2}{4} )$)
and $N_F (N_S)$ is the number of Weyl fermions (complex scalars) in
representation $R$. For N = 1 supersymmetric theories, Equation
\ref{eq:RGgeneral} can be made more compact.   We have \be
\label{eq:RGsusy} b_i = 3 C_2(G_i) - T_R \ N_\chi  \ee where the
first term takes into account the vector multiplets and $N_\chi$ is
the number of left-handed chiral multiplets in the representation
$R$ \cite{drw,2loop}.  The solution to the one loop RG equations is
given by \be \alpha_i(M_Z)^{-1} = \alpha_G^{-1} - \frac{b_i}{2 \pi}
\ln(\frac{M_G}{M_Z}) . \ee

For the SM we find \bea & {\bf b}_{SM} \equiv (b_1, b_2, b_3) & \\ &
= (-\frac{4}{3} N_{fam} - \frac{1}{10} N_H, \frac{22}{3} -
\frac{4}{3} N_{fam} - \frac{1}{6} N_H, 11 - \frac{4}{3} N_{fam}) &
\nn \eea where $N_{fam} (N_H)$ is the number of families (Higgs
doublets). For SUSY we have \be {\bf b}_{SUSY}  = \ee \be = (-2
N_{fam} - \frac{3}{5} N_{(H_u+H_d)}, 6 - 2 N_{fam} - N_{(H_u+H_d)},
9 - 2 N_{fam} ) \nn \ee where $N_{(H_u+H_d)}$ is the number of pairs
of Higgs doublets. Thus for the MSSM we have  \be {\bf b}^{MSSM} =
(- 33/5, -1, 3) . \ee

The one loop equations can be solved for the value of the GUT scale
$M_G$ and $\alpha_G$ in terms of the values of $\alpha_{EM}(M_Z)$
and $\sin^2\theta_W(M_Z)$.  We have (without including weak scale
threshold corrections)  \be \alpha_2(M_Z) =
\frac{\alpha_{EM}(M_Z)}{\sin^2\theta_W(M_Z)}, \; \alpha_1(M_Z) =
\frac{\frac{5}{3} \alpha_{EM}(M_Z)}{\cos^2\theta_W(M_Z)}  \ee and we
find  \be (\frac{3}{5} -\frac{8}{5} \ \sin^2\theta_W(M_Z))
\alpha_{EM}(M_Z)^{-1} = (\frac{b^{MSSM}_2 - b^{MSSM}_1}{2 \pi}) \ln
(\frac{M_G}{M_Z}) \ee which we use to solve for $M_G$.  Then we use
\be  \alpha_G^{-1} = \sin^2\theta_W(M_Z) \ \alpha_{EM}(M_Z)^{-1} +
\frac{b^{MSSM}_2}{2 \pi} \ln (\frac{M_G}{M_Z}) \ee to solve for
$\alpha_G$. We can then predict the value for the strong coupling
using \be \alpha_3(M_Z)^{-1} = \alpha_G^{-1} - \frac{b^{MSSM}_3}{2
\pi} \ln (\frac{M_G}{M_Z}). \ee

Given the experimental values  $\sin^2\theta_W(M_Z) \approx .23$ and
$\alpha_{EM}(M_Z)^{-1} \approx 128$ we find $M_G \approx 1.3 \times
10^{13}$ GeV with $N_H =1$ and $\alpha_G^{-1} \approx 42$ for the SM
with the one loop prediction for $\alpha_3(M_Z) \approx 0.07$. On
the other hand, for SUSY we find $M_G \approx 2.7 \times 10^{16}$
GeV,  $\alpha_G^{-1} \approx 24$ and the predicted strong coupling
$\alpha_3(M_Z) \approx 0.12$.  How well does this agree with the
data?  According to the PDG the average value of $\alpha_s(M_Z) =
0.1176 \pm 0.002$ \cite{pdg}.  So at one loop the MSSM is quite
good, while non-SUSY GUTs are clearly excluded.

At the present date,  the MSSM is compared to the data using 2 loop
RG running from the weak to the GUT scale with one loop threshold
corrections included at the weak scale.  These latter corrections
have small contributions from integrating out the W, Z, and top
quark. But the major contribution comes from integrating out the
presumed SUSY spectrum.  With a ``typical" SUSY spectrum and
assuming no threshold corrections at the GUT scale, one finds a
value for $\alpha_s(M_Z) \ge 0.127$ which is too large
\cite{Langacker:1995fk}.   It is easy to see where this comes from
using the approximate analytic formula \be \alpha_i^{-1}(M_Z) =
\alpha_G^{-1} - \frac{b^{MSSM}_i}{2 \pi} \ln(\frac{M_G}{M_Z}) +
\delta_i \ee where \be \delta_i = \delta_i^h + \delta_i^2 +
\delta_i^l . \ee The constants $\delta_i^2, \ \delta_i^l, \
\delta_i^h$ represent the 2 loop running effects \cite{2loop}, the
weak scale threshold corrections and the GUT scale threshold
corrections, respectively.  We have \be \delta_i^2 \approx -
\frac{1}{\pi} \sum_{j=1}^3  \frac{b^{MSSM}_{i j}}{b^{MSSM}_j} \log
\left[ 1 - b^{MSSM}_j \left( \frac{3 - 8 \sin^2\theta_W}{36
\sin^2\theta_W - 3} \right) \right] \ee where the matrix $
b^{MSSM}_{i j}$ is given by \cite{2loop} \be b^{MSSM}_{i j} = \left(
\begin{array}{ccc} \frac{199}{100} & \frac{27}{20} & \frac{22}{5} \\
 \frac{9}{20} & \frac{25}{4} & 6 \\ \frac{11}{20} & \frac{9}{4} &
 \frac{7}{2} \end{array} \right).  \ee   The light thresholds are
 given by \be \delta_i^l = \frac{1}{\pi} \sum_j {b^{l}_i}(j) \log
 (\frac{m_j}{M_Z} ) \ee where the sum runs over all states at the
 weak scale including the top, $W$, Higgs and the supersymmetric
 spectrum.  Finally the GUT scale threshold correction is given by
 \be \delta_i^h = - \frac{1}{2 \pi} \sum_\zeta b_i^\zeta \log (\frac{M_\zeta}{M_G}) . \ee

 In general the prediction for $\alpha_3(M_Z)$ is given by \bea
 \alpha_3^{-1}(M_Z)= &  (\frac{b_3-b_1}{b_2-b_1}) \alpha_2^{-1}(M_Z) -
 (\frac{b_3-b_2}{b_2-b_1}) \alpha_1^{-1}(M_Z) & \nn \\ & +
 (\frac{b_3-b_2}{b_2-b_1}) \delta_1 - (\frac{b_3-b_1}{b_2-b_1})
 \delta_2 + \delta_3 &  \nn \\ =  &  \frac{12}{7} \alpha_2^{-1}(M_Z) -
 \frac{5}{7} \alpha_1^{-1}(M_Z)& \nn \\ & + \frac{1}{7} (5 \delta_1 - 12
 \delta_2 + 7 \delta_3) & \nn \\ &  \equiv (\alpha_3^{LO})^{-1} + \delta_s  &  \eea
 where  $b_i \equiv b^{MSSM}_i$,  $\alpha_3^{LO}(M_Z)$ is the leading order one-loop result and $\delta_s \equiv  \frac{1}{7} (5 \delta_1 - 12
 \delta_2 + 7 \delta_3)$.  We find $\delta_s^2 \approx -0.82$ (Ref. \cite{Alciati:2005ur}) and
 $\delta_s^l = -0.04 + \frac{19}{28 \pi} \ln( \frac{T_{SUSY}}{M_Z})$
 where the first term takes into account the contribution of the
 $W$, top and the correction from switching from the $\overline{MS}$
 to $\overline{DR}$ RG schemes and
 (following Ref.  \cite{Carena:1993ag})
 \be \label{eq:Tsusy} T_{SUSY} = m_{\tilde H} (\frac{m_{\tilde W}}{m_{\tilde g}})^{28/19} \left[ (\frac{m_{\tilde l}}{m_{\tilde q}})^{3/19}
 (\frac{m_{H}}{m_{\tilde H}})^{3/19} (\frac{m_{\tilde W}}{m_{\tilde H}})^{4/19} \right]
 .\ee  For a Higgsino mass $m_{\tilde H} = 400$ GeV, a Wino mass $m_{\tilde W} = 300$ GeV, a gluino mass $m_{\tilde g} = 900$
 GeV and all other mass ratios of order one, we find $\delta_s^l
 \approx -0.12$.  If we assume $\delta_s^h = 0$, we find the
 predicted value of $\alpha_3(M_Z) = 0.135$.  In order to obtain a
 reasonable value of $\alpha_3(M_Z)$ with only weak scale threshold
 corrections, we need
$\delta_s^2 + \delta_s^l \approx 0$ corresponding to a value of
$T_{SUSY} \sim 5$ TeV.   But this is very difficult considering the
weak dependence $T_{SUSY}$ (Eqn. \ref{eq:Tsusy}) has on squark and
slepton masses. Thus in order to have $\delta_s \approx 0$ we need a
GUT scale threshold correction \be \label{eq:deltah} \delta_s^h
\approx + 0.94 . \ee

At the GUT scale we have \be \alpha_i^{-1}(M_G) = \alpha_G^{-1} +
\delta_i^h . \ee  Define \be  \tilde \alpha_G^{-1} = \frac{1}{7} [12
\alpha_2^{-1}(M_G) - 5 \alpha_1^{-1}(M_G) ] \ee (or if the GUT scale
is defined at the point where $\alpha_1$ and $\alpha_2$ intersect,
then $\tilde \alpha_G \equiv \alpha_1(M_G) = \alpha_2(M_G)$.  Hence,
in order to fit the data, we need a GUT threshold correction \be
\epsilon_3 \equiv \frac{\alpha_3(M_G) - \tilde \alpha_G}{\tilde
\alpha_G} = - \tilde \alpha_G \ \delta_s^h \approx  - 4 \%. \ee

\subsection{Nucleon Decay}

Baryon number is necessarily violated in any GUT~\cite{grs}.  In
$SU(5)$ nucleons decay via the exchange of gauge bosons with GUT
scale masses, resulting in dimension 6 baryon number violating
operators suppressed by $(1/M_G^2)$.  The nucleon lifetime is
calculable and given by \\ $\tau_N \propto M_G^4/(\alpha_G^2 \;
m_p^5)$.  The dominant decay mode of the proton (and the baryon
violating decay mode of the neutron), via gauge exchange, is $p
\rightarrow e^+ \; \pi^0$ ($n \rightarrow e^+ \; \pi^-$).  In any
simple gauge symmetry, with one universal GUT coupling and scale
($\alpha_G, \; M_G$), the nucleon lifetime from gauge exchange is
calculable.  Hence, the GUT scale may be directly observed via the
extremely rare decay of the nucleon.   In SUSY GUTs, the GUT scale
is of order $3\times 10^{16}$ GeV, as compared to the GUT scale in
non-SUSY GUTs which is of order $10^{15}$ GeV. Hence the dimension 6
baryon violating operators are significantly suppressed in SUSY
GUTs~\cite{drw} with $\tau_p \sim 10^{34 - 38}$ yrs.

However, in SUSY GUTs there are additional sources for baryon number
violation -- dimension 4 and 5 operators~\cite{bviol}.   Although
our notation does not change,  when discussing SUSY GUTs all fields
are implicitly bosonic superfields and the operators considered are
the so-called F terms which contain two fermionic components and the
rest scalars or products of scalars. Within the context of $SU(5)$
the dimension 4 and 5 operators have the form $({\bf 10 \; \bar 5 \;
\bar 5})$ $\supset  (U^c \; D^c \; D^c)  +  (Q \; L \; D^c)  + (E^c
\; L \; L)$ and $({\bf 10 \; 10 \; 10 \; \bar 5})$ $\supset (Q \;
Q\; Q\; L)  +  (U^c \; U^c\;  D^c\; E^c) \; +  $ $B$ and $L$
conserving terms, respectively.   The dimension 4 operators are
renormalizable with dimensionless couplings; similar to Yukawa
couplings.   On the other hand, the dimension 5 operators have a
dimensionful coupling of order ($1/M_G$).

The dimension 4 operators violate baryon number or lepton number,
respectively, but not both.  The nucleon lifetime is extremely short
if both types of dimension 4 operators are present in the low energy
theory.  However both types can be eliminated by requiring R parity.
In $SU(5)$ the Higgs doublets reside in a ${\bf 5_H,\; \bar 5_H}$
and R parity distinguishes the ${\bf \bar 5}$ (quarks and leptons)
from ${\bf \bar 5_H}$ (Higgs).  R parity~\cite{rparity} (or its
cousin, family reflection symmetry (see Dimopoulos and
Georgi~\cite{drw}and DRW ~\cite{drw2}) takes  $F \rightarrow -F, \;
H \rightarrow H$ with $F = \{ {\bf 10,\;  \bar 5} \}, \; H = \{ {\bf
\bar 5_H,\; 5_H} \}$. This forbids the dimension 4 operator $({\bf
10 \; \bar 5 \; \bar 5})$, but allows the Yukawa couplings of the
form $({\bf 10 \; \bar 5 \; \bar 5_H})$ and $({\bf 10 \; 10 \;
5_H})$.   It also forbids the dimension 3, lepton number violating,
operator $({\bf \bar 5 \; 5_H})$ $\supset (L \; H_u)$ with a
coefficient with dimensions of mass which, like the $\mu$ parameter,
could be of order the weak scale and the dimension 5, baryon number
violating, operator $({\bf 10 \; 10 \; 10 \; \bar 5_H})$ $\supset (Q
\; Q\; Q\; H_d) + \cdots$.

Note, in the MSSM it is possible to retain R parity violating
operators at low energy as long as they violate either baryon number
or lepton number only but not both.  Such schemes are natural if one
assumes a low energy symmetry, such as lepton number, baryon number
or a baryon parity~\cite{ir}.    However these symmetries cannot be
embedded in a GUT. Thus, in a SUSY GUT, only R parity can prevent
unwanted dimension four operators.  Hence, by naturalness arguments,
R parity must be a symmetry in the effective low energy theory of
any SUSY GUT.  This does not mean to say that R parity is guaranteed
to be satisfied in any GUT.

Note also, R parity distinguishes Higgs multiplets from ordinary
families.  In $SU(5)$, Higgs and quark/lepton multiplets have
identical quantum numbers; while in $E(6)$, Higgs and families are
unified within the fundamental ${\bf 27}$ representation. Only in
SO(10) are Higgs and ordinary families distinguished by their gauge
quantum numbers.   Moreover the $\mathbb{Z}_4$ center of $SO(10)$
distinguishes ${\bf 10}$s from ${\bf 16}$s and can be associated
with R parity~\cite{senjanovic}.

Dimension 5 baryon number violating operators may be forbidden at
tree level by symmetries in $SU(5)$, etc. These symmetries are
typically broken however by the VEVs responsible for the color
triplet Higgs masses. Consequently these dimension 5 operators are
generically generated via color triplet Higgsino exchange.  Hence,
the color triplet partners of Higgs doublets must necessarily obtain
mass of order the GUT scale.  The dominant decay modes from
dimension 5 operators are $p \rightarrow K^+ \; \bar \nu \;\; (n
\rightarrow K^0 \; \bar \nu)$. This is due to a simple symmetry
argument;  the operators $(Q_i \; Q_j\; Q_k\; L_l), \;\; (U^c_i \;
U^c_j\; D^c_k\; E^c_l)$ (where $i,\; j,\; k,\; l = 1,2,3$ are family
indices and color and weak indices are implicit) must be invariant
under $SU(3)_C$ and $SU(2)_L$.  As a result their color and weak
doublet indices must be anti-symmetrized.  However since these
operators are given by bosonic superfields, they must be totally
symmetric under interchange of all indices. Thus the first operator
vanishes for $i = j = k$ and the second vanishes for $i = j$.  Hence
a second or third generation particle must appear in the final
state~\cite{drw2}.

\begin{figure}
\includegraphics[height=.2\textheight]{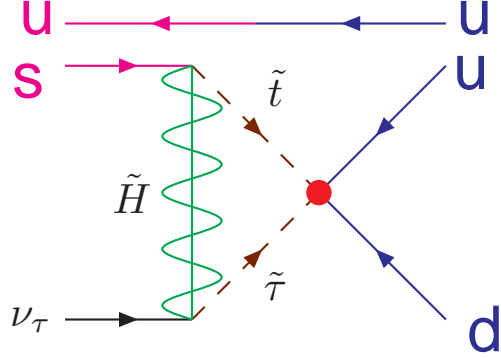}
\caption{\label{fig:pdecay} The effective four fermi operator for
proton decay obtained by integrating out sparticles at the weak
scale.}
\end{figure}

The dimension 5 operator contribution to proton decay requires a
sparticle loop at the SUSY scale to reproduce an effective dimension
6 four fermi operator for proton decay (see Fig. \ref{fig:pdecay}).
The loop factor is of the form  \be (LF) \propto \frac{\lambda_t \;
\lambda_\tau}{16 \pi^2} \frac{\sqrt{\mu^2 + M_{1/2}^2}}{m_{16}^2}
\ee leading to a decay amplitude  \be A(p \rightarrow K^+ \bar \nu)
\propto  \frac{c \ c}{M_T^{eff}} \ (\rm LF) . \ee   In any
predictive SUSY GUT, the coefficients $c$ are 3 $\times$ 3 matrices
related to (but not identical to) Yukawa matrices.  Thus these tend
to suppress the proton decay amplitude.   However this is typically
not sufficient to be consistent with the experimental bounds on the
proton lifetime.   Thus it is also necessary to minimize the loop
factor, (LF).   This can be accomplished by taking $\mu, M_{1/2}$
{\em small} and $m_{16}$ {\em large}.   Finally the effective Higgs
color triplet mass $M_T^{eff}$  must be MAXIMIZED.  With these
caveats, it is possible to obtain rough theoretical bounds on the
proton lifetime given by \cite{Lucas:1995ic,Altarelli:2000fu,dmr}
\be \tau_{p \rightarrow K^+ \bar \nu} \leq ( \frac{1}{3} - 3 )
\times 10^{34} \; {\rm yrs.} . \ee

\subsection{Gauge Coupling Unification and Proton Decay}

\begin{figure}
\includegraphics[height=.1\textheight]{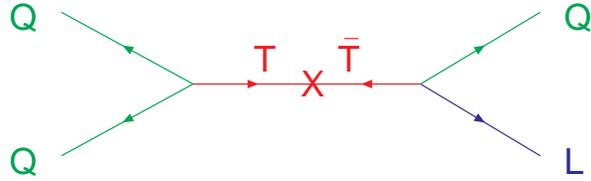}
\caption{\label{fig:dim5op} The effective dimension 5 operator for
proton decay.}
\end{figure}

The dimension 5 operator (see Fig. \ref{fig:dim5op}) is given in
terms of the matrices $c$ and an effective Higgs triplet mass by \be
\frac{1}{M^{eff}_T} \left[ \ Q \ \frac{1}{2} \ c_{qq} Q \ Q \ c_{ql}
L + \overline U \ c_{ud} \overline D \ \overline U \ c_{ue}
\overline E \ \right] . \ee Note, $M^{eff}_T$ can be much greater
than $M_G$ without fine-tuning and without having any particle with
mass greater than the GUT scale. Consider a theory with two pairs of
Higgs $5_i$ and $\bar 5_i$ with $i = 1,2$ at the GUT scale with only
$5_1, \ \bar 5_1$ coupling to quarks and leptons. Then we have \be
\frac{1}{M^{eff}_T} = ( M_T^{-1} )_{11} . \ee

If the Higgs color triplet mass matrix is given by  \be M_T = \left(
\begin{array}{cc} 0 &  M_G \\  M_G & X
\end{array} \right) \ee then we have \be \frac{1}{M^{eff}_T} \equiv
\frac{X}{M_G^2}. \ee  Thus for $X << M_G$ we obtain $M^{eff}_T
>> M_G$.

We assume that the Higgs doublet mass matrix, on the other hand, is
of the form \be M_D = \left(
\begin{array}{cc} 0 &  0 \\  0 & X
\end{array} \right) \ee  with two light Higgs doublets.   Note this
mechanism is natural in $S0(10)$
\cite{Dimopoulos:1981xm,Babu:1993we} with a superpotential of the
form   \be W \supset 10 \ 45 \ 10^\prime + X \ (10^\prime)^2 \ee
with only $10$ coupling to quarks and leptons, $X$ is a gauge
singlet and $\langle 45 \rangle = (B-L) \ M_G$.

Recall $\epsilon_3 \equiv \frac{(\alpha_3(M_G) - \tilde
\alpha_G)}{\tilde \alpha_G} \sim - 4\%$.   At one loop we find \be
\epsilon_3 = \epsilon_3^{\rm Higgs} + \epsilon_3^{\rm GUT \;
breaking} + \cdots . \ee  Moreover \be \epsilon_3^{\rm Higgs} =
\frac{3 \alpha_G}{5 \pi} \ln(\frac{M^{eff}_T}{M_G}) . \ee  See Table
\ref{tab:ep3} for the contribution to $\epsilon_3$ in Minimal SUSY
$SU(5)$, and in an $SU(5)$ and $SO(10)$ model with natural Higgs
doublet-triplet splitting.

\begin{table} \caption{Contribution to $\epsilon_3$ in three different GUT
models.} \label{tab:ep3}
$$
\begin{array}{|l|c|c|c|} \hline
{\rm Model}  &   {\rm Minimal}  & {SU_5}   & {\rm Minimal}  \\
& SU_5 & {\rm ``Natural" D/T} \mbox{\protect\cite{Altarelli:2000fu}} & SO_{10} \mbox{\protect\cite{dmr}} \\
\hline
\epsilon_3^{\rm GUT breaking} &  0  &  - 7.7 \%  &  -10 \% \\
\epsilon_3^{\rm Higgs} &  - 4 \% & + 3.7 \% &  + 6 \% \\
\hline M^{eff}_T {\rm [GeV]} &  2 \times 10^{14} & 3 \times 10^{18}
& 6 \times
10^{19} \\
\hline
\end{array}$$
\end{table}

Recent Super-Kamiokande bounds on the proton lifetime severely
constrain these dimension 6 and 5 operators with $\tau_{(p
\rightarrow e^+ \pi^0)} > 8.0 \times 10^{33}$ yrs \cite{susy08}, and
$\tau_{(p \rightarrow  K^+ \bar \nu)} > 2.3 \times 10^{33}$ yrs (92
ktyr) at (90\% CL) based on the listed exposures~\cite{superk}.
These constraints are now sufficient to rule out minimal SUSY
$SU(5)$~\cite{murayama}.  The upper bound on the proton lifetime
from these theories (particularly from dimension 5 operators) is
approximately a factor of 5 above the experimental bounds. These
theories are also being pushed to their theoretical limits.  Hence
if SUSY GUTs are correct, then nucleon decay must be seen soon.

\subsection{Yukawa coupling unification}
\subsubsection{3rd generation, $b - \tau$  or $t - b - \tau$ unification}

In $SU(5)$, there are two independent renormalizable Yukawa
interactions given by  $\lambda_t \; ({\bf 10 \; 10 \;  5_H}) $ $+
\; \lambda \; ({\bf 10 \; \bar 5 \; \bar 5_H})$.   These contain the
SM interactions $\lambda_t \; ({\bf Q \; u^c \; H_u})$  $ + \;
\lambda \; ({\bf Q \; d^c \;  H_d} \; + \; {\bf e^c \; L \; H_d})$.
Hence, at the GUT scale we have the tree level relation,  $\lambda_b
= \lambda_\tau \equiv \lambda$~\cite{chanowitz}.   In $SO(10)$ (or
Pati-Salam) there is only one independent renormalizable Yukawa
interaction given by $\lambda \; ({\bf 16 \; 16 \;  10_H})$ which
gives the tree level relation, $\lambda_t = \lambda_b = \lambda_\tau
\equiv \lambda$~\cite{so10yuk,hrr,so10yuksusy}. Note, in the
discussion above we assume the minimal Higgs content with Higgs in
${\bf 5,\; \bar 5}$ for $SU(5)$ and ${\bf 10}$ for $SO(10)$.  With
Higgs in higher dimensional representations there are more possible
Yukawa couplings~\cite{Lazarides:1980nt,Bajc:2002iw,Goh:2003sy}.

In order to make contact with the data, one now renormalizes the
top, bottom and $\tau$ Yukawa couplings, using two loop RG
equations, from $M_G$ to $M_Z$.   One then obtains the running quark
masses  $m_t(M_Z)\; = \;\lambda_t(M_Z)\; v_u$, $\;\; m_b(M_Z)\;  =\;
\lambda_b(M_Z)\; v_d$ and   $\; m_\tau(M_Z)\; =\;
\lambda_\tau(M_Z)\; v_d$ where $<H_u^0> \equiv v_u = \sin\beta \;
v/\sqrt{2}$, $<H_d^0> \equiv v_d = \cos\beta\; v/\sqrt{2}$, $v_u/v_d
\equiv \tan\beta$ and  $v \sim 246$ GeV is fixed by the Fermi
constant, $G_{\mu}$.

Including one loop threshold corrections at $M_Z$ and additional RG
running, one finds the top, bottom and $\tau$ pole masses. In SUSY,
$b - \tau$ unification has two possible solutions with $\tan\beta
\sim 1$ or $40 - 50$. The small $\tan\beta$ solution is now
disfavored by the LEP limit, $\tan\beta >
2.4$~\cite{lep}.\footnote{However, this bound disappears if one
takes $M_{SUSY} = 2$ TeV and $m_t = 180$ GeV~\cite{Carena:2002es}.}
The large $\tan\beta$ limit overlaps the $SO(10)$ symmetry relation.

When  $\tan\beta$ is large there are significant weak scale
threshold corrections to down quark and charged lepton masses from
either gluino and/or chargino loops~\cite{yukawacorr}. Yukawa
unification (consistent with low energy data) is only possible in a
restricted region of SUSY parameter space with important
consequences for SUSY searches~\cite{bdr}.

Consider a minimal $SO_{10}$ SUSY model [MSO$_{10}$SM]~\cite{bdr}.
Quarks and leptons of one family reside in the $\bf 16$ dimensional
representation, while the two Higgs doublets of the MSSM reside in
one $\bf 10$ dimensional representation.  For the third generation
we assume the minimal Yukawa coupling term given by $ {\bf \lambda \
16 \ 10 \ 16 }. $ On the other hand, for the first two generations
and for their mixing with the third, we assume a hierarchical mass
matrix structure due to effective higher dimensional operators.
Hence the third generation Yukawa couplings satisfy $\lambda_t =
\lambda_b = \lambda_\tau = \lambda_{\nu_\tau} = {\bf \lambda}$.

Soft SUSY breaking parameters are also consistent with $SO_{10}$
with (1) a universal gaugino mass $M_{1/2}$, (2) a universal squark
and slepton mass $m_{16}$,\footnote{$SO_{10}$ does not require all
sfermions to have the same mass. This however may be enforced by
non--abelian family symmetries or possibly by the SUSY breaking
mechanism.} (3) a universal scalar Higgs mass $m_{10}$, and (4) a
universal A parameter $A_0$. In addition we have the supersymmetric
(soft SUSY breaking) Higgs mass parameters $\mu$ ($B \mu$).  $B \mu$
may, as in the CMSSM, be exchanged for $\tan\beta$. Note, not all of
these parameters are independent. Indeed, in order to fit the low
energy electroweak data, including the third generation fermion
masses, it has been shown that $A_0, \ m_{10}, \ m_{16}$ must
satisfy the constraints~\cite{bdr}
\bea A_0 \approx - 2 \ m_{16}; & m_{10} \approx \sqrt{2} \ m_{16}  %\nonumber
\label{eq:constraint1}
\\
m_{16} > 1.2 \; {\rm TeV}; & \mu, \ M_{1/2} \ll m_{16} %\nonumber
\label{eq:constraint2} \eea with \be \tan\beta \approx 50.
\label{eq:tanbeta} \ee
This result has been confirmed by several independent
analyses~\cite{Tobe:2003bc,Auto:2003ys,Baer:2008jn}.\footnote{Note,
different regions of parameter space consistent with Yukawa
unification have also been discussed
in~\cite{Tobe:2003bc,Auto:2003ys,Balazs:2003mm}.} Although the
conditions (Eqns.~\ref{eq:constraint1}, \ref{eq:constraint2}) are
not obvious, it is however easy to see that
(Eqn.~(\ref{eq:tanbeta})) is simply a consequence of third
generation Yukawa unification, since $m_t(m_t)/m_b(m_t) \sim
\tan\beta$.

Finally, as a bonus, these same values of soft SUSY breaking
parameters, with $m_{16} \gg$ TeV, result in two very interesting
consequences.  Firstly, it ``naturally" produces an inverted scalar
mass hierarchy [ISMH]~\cite{scrunching}. With an ISMH, squarks and
sleptons of the first two generations obtain mass of order $m_{16}$
at $M_Z$. The stop, sbottom, and stau, on the other hand, have mass
less than (or of order) a TeV. An ISMH has two virtues. (1) It
preserves ``naturalness" (for values of $m_{16}$ which are not too
large), since only the third generation squarks and sleptons couple
strongly to the Higgs. (2) It ameliorates the SUSY CP and flavor
problems, since these constraints on CP violating angles or flavor
violating squark and slepton masses are strongest for the first two
generations, yet they are suppressed as $1/m_{16}^{2}$.  For $m_{16}
> $ a few TeV, these constraints are
weakened~\cite{masieroetal}.  Secondly, Super--Kamiokande bounds on
$\tau(p \rightarrow K^+ \bar \nu) \ > 2.3 \times 10^{33}$
yrs~\cite{superk} constrain the contribution of dimension 5 baryon
and lepton number violating operators. These are however minimized
with $\mu, \ M_{1/2} \ll m_{16}$~\cite{dmr}.

\subsubsection{Three families}

Simple Yukawa unification is not possible for the first two
generations of quarks and leptons.  Consider the $SU(5)$ GUT scale
relation $\lambda_b = \lambda_\tau$.  If extended to the first two
generations one would have $\lambda_s = \lambda_\mu$, $\lambda_d =
\lambda_e$ which gives $\lambda_s/\lambda_d =
\lambda_\mu/\lambda_e$. The last relation is a renormalization group
invariant and is thus satisfied at any scale. In particular, at the
weak scale one obtains $m_s/m_d = m_\mu/m_e$ which is in serious
disagreement with the data with $m_s/m_d \sim 20$ and $m_\mu/m_e
\sim 200$.  An elegant solution to this problem was given by Georgi
and Jarlskog~\cite{gj}. Of course, a three family model must also
give the observed CKM mixing in the quark sector. Note, although
there are typically many more parameters in the GUT theory above
$M_G$, it is possible to obtain effective low energy theories with
many fewer parameters making strong predictions for quark and lepton
masses.

It is important to note that grand unification alone is not
sufficient to obtain predictive theories of fermion masses and
mixing angles.  Other ingredients are needed.  In one approach
additional global family symmetries are introduced (non-abelian
family symmetries can significantly reduce the number of arbitrary
parameters in the Yukawa matrices).  These family symmetries
constrain the set of effective higher dimensional fermion mass
operators.  In addition, sequential breaking of the family symmetry
is correlated with the hierarchy of fermion masses. Three-family
models exist which fit all the data, including neutrino masses and
mixing~\cite{guts}.  In a completely separate approach for $SO(10)$
models, the Standard Model Higgs bosons are contained in the higher
dimensional Higgs representations including the {\bf 10},
$\overline{\bf 126}$ and/or {\bf 120}.   Such theories have been
shown to make predictions for neutrino masses and mixing
angles~\cite{Lazarides:1980nt,Bajc:2002iw,Goh:2003sy}.   Some simple
patterns of fermion masses (see Table \ref{tab:patterns}) must be
incorporated into any successful model.

\begin{table} \caption{Patterns of Masses and Mixing}
\begin{tabular}{|ll|}  \hline
$\lambda_t = \lambda_b = \lambda_\tau = \lambda_{\nu_\tau}$ &
$SO(10) @ M_G$  \\
\hline $ \lambda_s \sim \frac{1}{3} \lambda_\mu , \;\; \lambda_d
\sim 3 \lambda_e $ & $@ M_G $ \mbox{\protect\cite{gj,hrr}} \\
$ m_s \approx 4 \cdot \frac{1}{3} m_\mu ,  \;\; m_d \approx 4 \cdot
3 m_e $ & $ @ M_Z $ \\ \hline $ \lambda_d \lambda_s \lambda_b
\approx
\lambda_e \lambda_\mu \lambda_\tau$ & $SU(5) @ M_G$  \\
$ Det(m_d) \approx  Det(m_e)$ & $ @ M_G $ \\ \hline $ V_{us} \approx
(\sqrt{m_d/m_s} - i \sqrt{m_u/m_c})$ & \mbox{\protect\cite{fritzsch,Kim:2004ki}}  \\
$V_{ub}/V_{cb} \approx \sqrt{m_u/m_c}$ & \mbox{\protect\cite{Hall:1993ni}} \\
$ V_{cb}  \sim  m_s/m_b \sim \sqrt{m_c/m_t}$ &
\mbox{\protect\cite{hrr}}
\\ \hline
\end{tabular} \label{tab:patterns}
\end{table}

\subsection{Neutrino Masses}

Atmospheric and solar neutrino oscillations require neutrino masses.
Adding three ``sterile" neutrinos $\nu^c$ with the Yukawa coupling
$\lambda_\nu \; ({\bf \nu^c \; L \; H_u})$, one easily obtains three
massive Dirac neutrinos with mass $m_\nu = \lambda_\nu \;
v_u$.\footnote{Note, these ``sterile" neutrinos are quite naturally
identified with the right-handed neutrinos necessarily contained in
complete families of $SO(10)$ or Pati-Salam.} However in order to
obtain a tau neutrino with mass of order $ 0.1\; {\rm eV} $, one
needs $\lambda_{\nu_\tau}/\lambda_\tau \leq 10^{-10}$. The see-saw
mechanism, on the other hand, can naturally explain such small
neutrino masses~\cite{minkowski,yanagida}. Since $\nu^c$ has no SM
quantum numbers, there is no symmetry (other than global lepton
number) which prevents the mass term $\frac{1}{2} \; \nu^c \; M \;
\nu^c$. Moreover one might expect $M \sim M_G$.   Heavy ``sterile"
neutrinos can be integrated out of the theory, defining an effective
low energy theory with only light active Majorana neutrinos with the
effective dimension 5 operator $\frac{1}{2} \; ({\bf L\;  H_u}) \;
\lambda_\nu^T \; M^{-1} \; \lambda_\nu \; ({\bf L\; H_u})$.  This
then leads to a $3 \times 3$ Majorana neutrino mass matrix ${\bf m}
= m_\nu^T \; M^{-1} \; m_\nu$.

Atmospheric neutrino oscillations require neutrino masses with
$\Delta m_\nu^2 \sim 3 \times 10^{-3}$ eV$^2$ with maximal mixing,
in the simplest two neutrino scenario.  With hierarchical neutrino
masses $m_{\nu_\tau} = \sqrt{\Delta m_\nu^2} \sim 0.055$ eV.
Moreover via the ``see-saw" mechanism $m_{\nu_\tau} = m_t(m_t)^2/(3
M)$. Hence one finds $M \sim 2 \times 10^{14}$ GeV;  remarkably
close to the GUT scale.  Note we have related the neutrino Yukawa
coupling to the top quark Yukawa coupling  $\lambda_{\nu_\tau} =
\lambda_t$ at $M_G$ as given in $SO(10)$ or $SU(4)\times SU(2)_L
\times SU(2)_R$.  However at low energies they are no longer equal
and we have estimated this RG effect by $\lambda_{\nu_\tau}(M_Z)
\approx \lambda_t(M_Z)/\sqrt{3}$.

\subsection{$SO(10)$ GUT with $[D_3 \times [U(1) \times \mz_2 \times \mz_3]]$ Family Symmetry}

A complete model for fermion masses was given in Refs.
\cite{Dermisek:2005ij,Dermisek:2006dc}.    Using a global $\chi^2$
analysis, it has been shown that the model fits all fermion masses
and mixing angles, including neutrinos, and  a minimal set of
precision electroweak observables.  The model is consistent with
lepton flavor violation and lepton electric dipole moment bounds. In
two recent papers, Ref. \cite{Albrecht:2007ii,Altmannshofer:2008vr},
the model was also tested by flavor violating processes in the B
system.

The model is an $SO(10)$ SUSY GUT with an additional $D_3 \times
[U(1) \times \mz_2 \times \mz_3]$ family symmetry. The symmetry
group fixes the following structure for the superpotential \be W =
W_f + W_\nu ~,  \ee with \bea \label{Wcf}
W_f =& \textbf{16}_3 \, \textbf{10}\, \textbf{16}_3+\textbf{16}_a \, \textbf{10}\, \chi_a & \\
&+\bar{\chi}_a (M_{\chi}\, \chi_a+\textbf{45}
\,\frac{\phi_a}{\hat{M}}\, \textbf{16}_3
+\textbf{45} \,\frac{\tilde{\phi}_a}{\hat{M}} \, \textbf{16}_a+A\, \textbf{16}_a)~, & \nn \\
W_\nu =&\ov{\textbf{16}} (\lambda_2\,
N_a\,\textbf{16}_a+\lambda_3\,N_3\,\textbf{16}_3) & \label{eq:wnu}
\\ & +\frac{1}{2} (S_a\,N_a\,N_a+S_3\,N_3\, N_3)~& . \nn \eea
The first two families of quarks and leptons are contained in the
superfield $\textbf{16}_a,\:a=1,2$, which transforms under
SO(10)$\times D_3$ as $(\textbf{16}, \textbf{$2_A$})$, whereas the
third family in $\textbf{16}_3$ transforms as $
(\textbf{16},\textbf{$1_B$})$. The two MSSM Higgs doublets $H_{u}$
and $H_d$ are contained in a $\textbf{10}$. As can be seen from the
first term on the right-hand side of (\ref{Wcf}), Yukawa unification
$\lambda_t=\lambda_b=\lambda_\tau=\lambda_{\nu_\tau}$ at $M_G$ is
obtained {\em only} for the third generation, which is directly
coupled to the Higgs $\textbf{10}$ representation. This immediately
implies large $\tan\beta \approx 50$ at low energies and constrains
soft SUSY breaking parameters.

The effective Yukawa couplings of the first and second generation
fermions are generated hierarchically via the Froggatt-Nielsen [FN]
mechanism \cite{froggatt} as follows. Additional fields are
introduced, i.e. the $\textbf{45}$ which is an adjoint of SO(10),
the SO(10) singlet flavon fields $\phi^a, \tilde {\phi^a}, A$ and
the Froggatt-Nielsen [FN] states $\chi_a, \bar{\chi}_a$. The latter
transform as a $(\textbf{16},\textbf{$2_A$})$ and a
$(\ov{\textbf{16}},\textbf{$2_A$})$, respectively, and receive
masses of O$(M_{G})$ as $M_\chi$ acquires an SO(10) breaking VEV.
Once they are integrated out, they give rise to effective mass
operators which, together with the VEVs of the flavon fields, create
the Yukawa couplings for the first two generations. This mechanism
breaks systematically the full flavor symmetry and produces the
right mass hierarchies among the fermions.

\begin{table}
\begin{tabular}{|lcc|}
\hline
Sector & \# & Parameters \\
\hline \hline
gauge & 3 & $\alpha_G$, $M_G$, $\epsilon_3$, \\
SUSY (GUT scale) & 5 & $m_{16}$, $M_{1/2}$, $A_0$, $m_{H_u}$, $m_{H_d}$, \\
textures & 11 & $\epsilon$, $\epsilon^\prime$, $\lambda$, $\rho$, $\sigma$, $\tilde \epsilon$, $\xi$, \\
neutrino & 3 & $M_{R_1}$, $M_{R_2}$, $M_{R_3}$, \\
SUSY (EW scale) & 2 & $\tan\beta$, $\mu$ \\
\hline \hline
\end{tabular}
\caption{The 24 parameters defined at the GUT scale which are used
to minimize $\chi^2$.} \label{tab:parameters}
\end{table}

Upon integrating out the FN states one obtains Yukawa matrices for
up-quarks, down-quarks, charged leptons and neutrinos given by \bea
&Y_u=\left(
\begin{array}{ccc}
0 & \varepsilon' \,\rho & -\varepsilon\,\xi \\
-\varepsilon' \, \rho & \tilde{\varepsilon} \,\rho & -\varepsilon \\
\varepsilon\,\xi & \varepsilon & 1
\end{array}
\right)\,\lambda~,&~~\nn \\ & Y_d=\left(
\begin{array}{ccc}
0 & \varepsilon'  & -\varepsilon\,\xi\,\sigma \\
-\varepsilon'  & \tilde{\varepsilon}  & -\varepsilon\,\sigma \\
\varepsilon\,\xi & \varepsilon & 1
\end{array}
\right)\,\lambda~,\nn \\
&Y_e=\left(
\begin{array}{ccc}
0 & -\varepsilon'  & 3\,\varepsilon\,\xi \\
\varepsilon'  & 3\,\tilde{\varepsilon}  & 3\,\varepsilon \\
-3\,\varepsilon\,\xi\,\sigma & -3\,\varepsilon\,\sigma & 1
\end{array}
\right)\,\lambda~,&~~ \nn \\& Y_\nu=\left(
\begin{array}{ccc}
0 & -\varepsilon' \,\omega & \frac{3}{2}\,\varepsilon\,\xi \,\omega\\
\varepsilon'  \,\omega& 3\,\tilde{\varepsilon}\,\omega &
\frac{3}{2}\,\varepsilon\,\omega \\ -3\,\varepsilon\,\xi\,\sigma &
-3\,\varepsilon\,\sigma & 1
\end{array}
\right)\,\lambda~. \label{Y-textures} \eea From eqs.
(\ref{Y-textures}) one can see that the flavor hierarchies in the
Yukawa couplings are encoded in terms of the four complex parameters
$\rho, \sigma, \tilde \varepsilon, \xi$ and the additional real ones
$\varepsilon, \varepsilon', \lambda$.

For neutrino masses one invokes the See-Saw
mechanism~\cite{minkowski,yanagida}. In particular, three SO(10)
singlet Majorana fermion fields $N_a, N_3$ $(a=1,2)$ are introduced
via the contribution of $\frac{1}{2}\, (S_a\,N_a \, N_a+S_3
\,N_3\,N_3)$ to the superpotential (Eqn. \ref{eq:wnu}). The mass
term $\frac{1}{2}\,N\,M_N\,N$ is produced when the flavon fields
acquire VEVs $\langle S_a\rangle=M_{N_a}$ and $\langle
S_3\rangle=M_{N_3}$. Together with a $\ov{\textbf{16}}$ Higgs one is
allowed to introduce the interaction terms $\ov{\textbf{16}}
\,(\lambda_2 \, N_a\, \textbf{16}_a+\lambda_3 \, N_3\,
\textbf{16}_3)$ (Eqn. \ref{eq:wnu}).  This then generates a mixing
matrix $V$ between the right-handed neutrinos and the additional
singlets ($\nu^c \,  V \, N$), when the $\ov{\textbf{16}}$ acquires
an SO(10) breaking VEV $\langle \ov{\textbf{16}} \rangle_{\nu^c} =
v_{16}$. The resulting effective right-handed neutrino mass terms
are given by \bea
 W_N= \nu^c \, V\,N+\frac{1}{2}\,N\,M_N\,N~,
\eea \bea V= & v_{16}\left(\begin{array}{ccc}
0 & \lambda_2 & 0 \\
\lambda_2 & 0 & 0 \\
0 & 0 & \lambda_3
\end{array}\right)~,~~~~ & \nn \\
M_N= & {\rm diag}(M_{N_1},M_{N_2},M_{N_3})~. \eea Diagonalization
leads to the effective right-handed neutrino Majorana mass \bea M_R
= - V \, M_N^{-1}\, V^T \equiv - {\rm diag}(M_{R_1},M_{R_2},M_{R_3})
~. \eea By integrating out the EW singlets $\nu^c$ and $N$, which
both receive GUT scale masses, one ends up with the light neutrino
mass matrix at the EW scale given by the usual see-saw formula \bea
\mc M = m_\nu\, M_R^{-1}\,m_\nu^T~. \eea

\begin{table}[]
\begin{tabular}{|lc||lc|}
\hline
Observable & Value($\sigma_{\rm exp}$)  & Observable & Value($\sigma_{\rm exp}$)  \\
\hline \hline
$M_W$ & $80.403(29)$ &  $M_\tau$ & $1.777(0)$  \\
$M_Z$ & $91.1876(21)$ &   $M_\mu$ & $0.10566(0)$ \\
$10^{5} G_\mu$ & $1.16637(1)$ &  $10^3 M_e$ & $0.511(0)$  \\
$1/\alpha_\text{em}$ & $137.036$  & $|V_{us}|$ & $0.2258(14)$ \\
$\alpha_s(M_Z)$ & $0.1176(20)$  & $10^3 |V_{ub}|$ & $4.1(0.4)$  \\
$M_t$ & $170.9(1.8)$  & $10^2 |V_{cb}|$ & $4.16(7)$ \\
$m_b(m_b)$ & $4.20(7)$  & $\sin 2 \beta$ & $0.675(26)$  \\
$m_c(m_c)$ & $1.25(9)$ & $10^3 \Delta m_{31}^2$ [eV$^2$]& $2.6(0.2)$  \\
$m_s(2~{\rm GeV})$ & $0.095(25)$  & $10^5 \Delta m_{21}^2$ [eV$^2$]& $7.90(0.28)$  \\
$m_d(2~{\rm GeV})$ & $0.005(2)$ & $\sin^2 2 \theta_{12}$ & $0.852(32)$  \\
$m_u(2~{\rm GeV})$ & $0.00225(75)$  & $\sin^2 2 \theta_{23}$ & $0.996(18)$ \\
\hline \hline
\end{tabular}
\caption{Flavor conserving observables used in the fit. Dimensionful
quantities are expressed in GeV, unless otherwise specified
\protect\cite{Albrecht:2007ii}.} \label{tab:obs}
\end{table}

\begin{table}[ht]
\begin{tabular}{|lc|}
\hline
Observable & Value($\sigma_{\rm exp}$)($\sigma_{\rm theo}$) \\
\hline \hline
$10^3 \epsilon_K$ & 2.229(10)(252) \\
$\Delta M_s / \Delta M_d$ & 35.0(0.4)(3.6)  \\
$10^4$ BR$(B \to X_s \gamma)$ & 3.55(26)(46)  \\
$10^6$ BR$(B \to X_s \ell^+ \ell^-)$ & 1.60(51)(40)  \\
$10^4$ BR$(B^+ \to \tau^+ \nu)$ & 1.31(48)(9)  \\
BR$(B_s \to \mu^+ \mu^-)$ & $< 1.0 \times 10^{-7}$ \\
\hline \hline
\end{tabular}
\caption{FC observables used in the fit
\protect\cite{Albrecht:2007ii}.} \label{tab:FCobs}
\end{table}

\begin{table}[r]
\begin{tabular}{|lc|}
\hline
Observable & Lower Bound  \\
\hline \hline
$M_{h_0}$ & $114.4$ GeV  \\
$m_{\tilde t}$ & $60$ GeV \\
$m_{\tilde\chi^+}$ & $104$ GeV  \\
$m_{\tilde g}$ & $195$ GeV \\
\hline \hline
\end{tabular}
\caption{Mass bounds used in the fit
\protect\cite{Albrecht:2007ii}.} \label{tab:bounds}
\end{table}

The model has a total of 24 arbitrary parameters, with all except
$\tan\beta$ defined at the GUT scale (see Table
\ref{tab:parameters}). Using a two loop RG analysis the theory is
redefined at the weak scale. Then a $\chi^2$ function is constructed
with low energy observables.  In Ref. \cite{Dermisek:2006dc} fermion
masses and mixing angles, a minimal set of precision electroweak
observables and the branching ratio BR($b \rightarrow s \gamma$)
were included in the $\chi^2$ function. Then predictions for lepton
flavor violation, lepton electric dipole moments, Higgs mass and
sparticle masses were obtained.  The $\chi^2$ fit was quite good.
The light Higgs mass was always around 120 GeV.  In the recent
paper, Ref. \cite{Albrecht:2007ii}, precision B physics observables
were added. See Tables \ref{tab:obs}, \ref{tab:FCobs} for the 28 low
energy observables and Table \ref{tab:bounds} for the 4 experimental
bounds included in their analysis.    The fits were not as good as
before with a minimum $\chi^2 \sim 25$ obtained for large values of
$m_{16} = 10$ TeV.

The dominant problem was due to constraints from the processes $B
\rightarrow X_s \gamma, \; B \rightarrow X_s \ell^+ \ell^-$.  The
latter process favors a coefficient $C_7$ for the operator \be O_7 =
 m_b \ \bar s_L \Sigma_{\mu \nu} b_R \ F^{\mu \nu}  \ee with $C_7 \sim
(C_7^{\rm SM})$,  while the former process only measures the
magnitude of $C_7$,  while the former process only measures the
magnitude of $C_7$. Note, the charged and neutral Higgs
contributions to ${\rm BR} (B\to X_s \gamma)$ are strictly positive.
While the sign of the chargino contribution, relative to the SM, is
ruled by the following relation \bea C_7^{\tilde \chi^+} \propto +
\mu A_t \tan\beta \times {\rm sign}(C_7^{\rm SM})~, \label{C7chi}
\eea with a positive proportionality factor, so it is opposite to
that of the SM one for $\mu > 0$ and $A_t < 0$.  Another problem was
$V_{ub}$ which was significantly smaller than present CKM fits.

In the recent analysis, Ref. \cite{Altmannshofer:2008vr},  it was
shown that better $\chi^2$ can be obtained by allowing for a 20\%
correction to Yukawa unification.  Note, this analysis only included
Yukawa couplings for the third family.  For a good fit, see Table
\ref{tab:fit-example}. We find $\tan\beta$ still large, $\tan\beta =
46$ and a light Higgs mass $m_h = 121$ GeV.   See Table
\ref{tab:fit-example} for the sparticle spectrum which should be
observable at the LHC.
\begin{table}[t]
\begin{tabular}{|lccc|}
\hline
Observable  &  Exp.  &  Fit  &  Pull  \\
\hline\hline
$M_W$  &  80.403  &  80.56  &  0.4  \\
$M_Z$  &  91.1876  &  90.73  &  \textbf{1.0}  \\
$10^{5}\; G_\mu$  &  1.16637  &  1.164  &  0.3  \\
$1/\alpha_\text{em}$  &  137.036  &  136.5  &  0.8  \\
$\alpha_s(M_Z)$  &  0.1176  &  0.1159  &  0.8  \\
$M_t$  &  170.9  &  171.3  &  0.2  \\
$m_b(m_b)$  &  4.20  &  4.28  &  \textbf{1.1}  \\
$M_\tau$  &  1.777  &  1.77  &  0.4  \\
$10^{4}\; \text{BR} (B \to X_s \gamma)$  &  3.55  &  2.72  &  \textbf{1.6}  \\
$10^{6}\; \text{BR} (B \to X_s \ell^+\ell^-)$  &  1.60  &  1.62  &  0.0  \\
$\Delta M_s / \Delta M_d$  &  35.05  &  32.4  &  0.7  \\
$10^{4}\; \text{BR} (B^+ \to \tau^+\nu)$  &  1.41  &  0.726  &  \textbf{1.4}  \\
$10^{8}\; \text{BR} (B_s \to \mu^+\mu^-)$  &  $<5.8 $ &  3.35  &  --  \\
\hline
\multicolumn{3}{|r}{total $\chi^2$:}  &  \textbf{8.78} \\
\hline
\end{tabular}
\hspace{2cm}
\begin{tabular}{|lc|lc|}
\hline
\multicolumn{2}{|l}{Input parameters} & \multicolumn{2}{|l|}{Mass predictions} \\
\hline\hline
$m_{16}$  &  $7000$  &  $M_{h^0}$  &  121.5  \\
$\mu$  &  $1369$  &  $M_{H^0}$  &  585  \\
$M_{1/2}$  &  $143$  &  $M_{A}$  &  586  \\
$A_0$  &  $-14301$  &  $M_{H^+}$  &  599  \\
$\tan\beta$  &  $46.1$  &  $m_{\tilde t_1}$  &  783  \\
$1/\alpha_G$  &  $24.7$  &  $m_{\tilde t_2}$  &  1728  \\
$M_G / 10^{16}$  &  $3.67$  &  $m_{\tilde b_1}$  &  1695  \\
$\epsilon_3 / \%$  &  $-4.91$  &  $m_{\tilde b_2}$  &  2378  \\
$(m_{H_u}/m_{16})^2$  &  $1.616$  &  $m_{\tilde \tau_1}$  &  3297  \\
$(m_{H_d}/m_{16})^2$  &  $1.638$  &  $m_{\tilde\chi^0_1}$  &  58.8  \\
$M_{R} / 10^{13}$  &  $8.27$  &  $m_{\tilde\chi^0_2}$  &  117.0  \\
$\lambda_u$  &  $0.608$  &  $m_{\tilde\chi^+_1}$  &  117.0  \\
$\lambda_d$  &  $0.515$  &  $M_{\tilde g}$  &  470  \\
\hline
\end{tabular}
\caption{Example of successful fit in the region with $b -\tau$
unification. Dimensionful quantities are expressed in powers of GeV.
Higgs, lightest stop and gluino masses are pole masses, while the
rest are running masses evaluated at $M_Z$
\cite{Altmannshofer:2008vr}.} \label{tab:fit-example}
\end{table} Finally, an analysis of dark matter for this model has been performed with
good fits to WMAP data \cite{Dermisek:2003vn}.\footnote{The authors
of Ref. \cite{Baer:2008jn} also analyze dark matter in the context
of the minimal $SO(10)$ model with Yukawa unification.   They have
difficulty fitting WMAP data.  We believe this is because they do
not adjust the CP odd Higgs mass to allow for dark matter
annihilation on the resonance.}

\subsection{Problems of 4D GUTs}

There are two aesthetic (perhaps more fundamental) problems
concerning 4d GUTs. They have to do with the complicated sectors
necessary for GUT symmetry breaking and Higgs doublet-triplet
splitting.   These sectors are sufficiently complicated that it is
difficult to imagine that they may be derived from a more
fundamental theory, such as string theory.  In order to resolve
these difficulties, it becomes natural to discuss grand unified
theories in higher spatial dimensions.  These are the so-called
orbifold GUT theories discussed in the next section.

Consider, for example, one of the simplest constructions in $SO(10)$
which accomplishes both tasks of GUT symmetry breaking and Higgs
doublet-triplet splitting \cite{Barr:1997hq}. Let there be a single
adjoint field, $A$, and {\it two} pairs of spinors, $C +
\overline{C}$ and $C' + \overline{C}'$. The complete Higgs
superpotential is assumed to have the form

\begin{equation}
W = W_A + W_C + W_{ACC'} + (T_1 A T_2 + S T_2^2).
\end{equation}

\noindent The precise forms of $W_A$ and $W_C$ do not matter, as
long as $W_A$ gives $\langle A \rangle$ the Dimopoulos-Wilczek form,
and $W_C$ makes the VEVs of $C$ and $\overline{C}$ point in the
$SU(5)$-singlet direction. For specificity we will take $W_A =
\frac{1}{4 M} {\rm tr} A^4 + \frac{1}{2} P_A ({\rm tr} A^2 + M_A^2)
+ f(P_A)$, where $P_A$ is a singlet, $f$ is an arbitrary polynomial,
and $M \sim M_G$. (It would be possible, also, to have simply $m \
{\rm Tr} A^2$, instead of the two terms containing $P_A$. However,
explicit mass terms for adjoint fields may be difficult to obtain in
string theory.) We take $W_C = X(\overline{C} C - P_C^2)$, where $X$
and $P_C$ are singlets, and $\langle P_C \rangle \sim M_G$.

The crucial term that couples the spinor and adjoint sectors
together has the form

\begin{equation}
W_{ACC'} = \overline{C}' \left( \left( \frac{P}{M_P} \right) A + Z
\right) C + \overline{C} \left( \left( \frac{\overline{P}}{M_P}
\right) A + \overline{Z} \right) C',
\end{equation}

\noindent where $Z$, $\overline{Z}$, $P$, and $\overline{P}$ are
singlets. $\langle P \rangle$ and $\langle \overline{P} \rangle$ are
assumed to be of order $M_G$. The critical point is that the VEVs of
the primed spinor fields will vanish, and therefore the terms in Eq.
(3) will not make a destabilizing contribution to $- F_A^* =
\partial W/\partial A$. This is the essence of the mechanism.

$W$ contains several singlets ($P_C$, $P$, $\overline{P}$, and $S$)
that are supposed to acquire VEVs of order $M_G$, but which are left
undetermined at tree-level by the terms so far written down. These
VEVs may arise radiatively when SUSY breaks, or may be fixed at tree
level by additional terms in $W$, possible forms for which will be
discussed below.

In $SU(5)$ the construction which gives natural Higgs
doublet-triplet splitting requires the $SU(5)$ representations ${\bf
75, \ 50, \ \overline{50}}$ and a superpotential of the form
\cite{missingpartner,Altarelli:2000fu}  \be W \supset 75^3 + M 75^2
+ 5_H \ 75 \ 50 + \bar 5_H \ 75 \ \overline{50} + 50 \ \overline{50}
\ X . \ee

\section{Orbifold GUTs}

\subsection{GUTs on a Circle}

As the first example of an orbifold GUT consider a pure $SO(3)$
gauge theory in 5 dimensions \cite{Dermisek:2002ri}.  The gauge
field is \be A_M \equiv A^a_M \ T^a,  \; a=1,2,3 ; \; M,N =
\{0,1,2,3,5\} .\ee The gauge field strength is given by \be F_{MN}
\equiv F^a_{MN} \ T^a =
\partial_M A_N - \partial_N A_M + i [A_M, A_N]  \ee where $T^a$ are
$SO(3)$ generators.  The Lagrangian is \be {\cal L}_5 = - \frac{1}{4
g_5^2 k} \ Tr (F_{MN} F^{MN}) \ee and we have $Tr (T^a \ T^b) \equiv
k \delta^{ab}$.  The inverse gauge coupling squared has mass
dimensions one.

Let us first compactify the theory on ${\cal M}_4 \times S^1$ with
coordinates  $\{ x_\mu, y \}$ and $y = [0, 2 \pi R)$. The theory is
invariant under the local gauge transformation \bea A_M(x_\mu, y) &
\rightarrow U \ A_M(x_\mu, y) \ U^\dagger - i U \ \partial_M \
U^\dagger, & \nn \\ & U = \exp( i \theta^a(x_\mu, y) \ T^a). & \eea

Consider the possibility $\partial_5 A_\mu \equiv 0$.   We have \be
F_{\mu 5} = \partial_\mu A_5 + i [A_\mu, A_5] \equiv  D_\mu \ A_5 .
\ee  We can then define \be \tilde \Phi \equiv A_5 \ \frac{\sqrt{2
\pi R}}{g_5} \equiv A_5/g \ee where $g_5 \equiv \sqrt{2 \pi R} \; g$
and $g$ is the dimensionless 4d gauge coupling.  The 5d Lagrangian
reduces to the Lagrangian for a 4d $SO(3)$ gauge theory with
massless scalar matter in the adjoint representation, i.e. \be {\cal
L}_5 = \frac{1}{2 \pi R} [ -\frac{1}{4 g^2 k} \ Tr(F_{\mu \nu}
F^{\mu \nu}) + \frac{1}{2 k} \ Tr(D_\mu \tilde \Phi \ D^\mu \tilde
\Phi) ] . \ee

In general we have the mode expansion \be A_M(x_\mu, y) = \sum_n [
a^n_M \cos n \frac{y}{R} + b^n_M \sin n \frac{y}{R} ] \ee where only
the cosine modes with $n = 0$ have zero mass.   Otherwise the 5d
Laplacian $\partial_M \partial^M = \partial_\mu \partial^\mu +
\partial_y \partial^y$ leads to Kaluza-Klein [KK] modes with
effective 4d mass \be m_n^2 = \frac{n^2}{R^2} . \ee

\subsection{Fermions in 5d}

The Dirac algebra in 5d is given in terms of the 4 $\times$ 4 gamma
matrices  $\gamma_M,  \; M = 0,1,2,3,5$ satisfying $\{ \gamma_M,
\gamma_N \} = 2 g_{MN}$.  A four component massless Dirac spinor
$\Psi(x_\mu, y)$ satisfies the Dirac equation \be i \gamma_M
\partial^M \Psi = 0 = i (\gamma_\mu \partial^\mu - \gamma_5
\partial_y) \Psi . \ee  In 4d the four
component Dirac spinor decomposes into two Weyl spinors with \be
\Psi = \left( \begin{array}{c} \psi_1 \\  i \sigma_2 \psi_2^*
\end{array} \right) = \left( \begin{array}{c} \psi_L \\  \psi_R
\end{array} \right)\ee where $\psi_{1,2}$ are two left-handed Weyl
spinors.   In general, we obtain the normal mode expansion for the
fifth direction given by \be  \psi_{L,R} =  \sum ( a_n(x) \cos n
\frac{y}{R} + b_n(x) \sin n \frac{y}{R} ). \ee If we couple this 5d
fermion to a local gauge theory, the theory is necessarily
vector-like; coupling identically to both $\psi_{L,R}$.

 We can obtain a chiral theory in 4d with the
following parity operation \be {\cal P} : \Psi(x_\mu, y) \rightarrow
\Psi(x_\mu, -y) = P \Psi(x_\mu, y) \ee  with $P =
-\gamma_5$.   We then have \bea  \Psi_L \sim \cos n \frac{y}{R} \nn \\
\Psi_R \sim \sin n \frac{y}{R}. \eea

\subsection{GUTs on an Orbi-Circle}

Let us briefly review the geometric picture of orbifold GUT models
compactified on an orbi-circle ${\rm S}^1/{\mathbb Z}_2$. The circle
$S^1 \equiv \mathbb{R}^1/{\cal T}$ where ${\cal T}$ is the action of
translations by $2 \pi R$.  All fields $\Phi$ are thus periodic
functions of $y$ (up to a finite gauge transformation), i.e. \be
{\cal T} : \Phi(x_\mu, y) \rightarrow \Phi(x_\mu, y + 2 \pi R) = T \
\Phi(x_\mu, y)  \ee where $T \in SO(3)$ satisfies $T^2 = 1$.  This
corresponds to the translation ${\cal T}$ being realized
non-trivially by a degree-2 Wilson line (i.e., background gauge
field - $\langle A_5 \rangle \neq 0$ with $T \equiv \exp (i \oint
\langle A_5 \rangle dy)$). Hence the space group of ${\rm
S}^1/\mz_2$ is composed of two actions, a translation, ${\cal
T}:\,y\rightarrow y+2\pi R$, and a space reversal, ${\cal
P}:\,y\rightarrow -y$. There are two (conjugacy) classes of fixed
points, $y=(2n)\pi R$ and $(2n+1)\pi R$, where $n\in\mz$.

\begin{figure}
\includegraphics[height=.04\textheight]{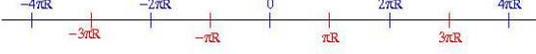}
\caption{\label{fig:epsart} The real line moded out by the space
group of translations, ${\cal T}$, and a $\mathbb{Z}_2$ parity,
${\cal P}$.}
\end{figure}

The space group multiplication rules imply ${\cal T}{\cal P}{\cal
T}={\cal P}$, so we can replace the translation by a composite
$\mz_2$ action ${\cal P}'={\cal P}{\cal T}: y\rightarrow -y+2\pi R$.
The orbicircle ${\rm S}^1/{\mathbb Z}_2$ is equivalent to an
${\mathbb R}/({\mathbb Z}_2\times{\mathbb Z}'_2)$ orbifold, whose
fundamental domain is the interval $[0,\,\pi R]$, and the two ends
$y=0$ and $y=\pi R$ are fixed points of the ${\mathbb Z}_2$ and
${\mathbb Z}_2'$ actions respectively.

A generic 5d field $\Phi$  has the following transformation
properties under the ${\mathbb Z}_2$ and ${\mathbb Z}'_2$
orbifoldings (the 4d space-time coordinates are suppressed),
\bea {\cal P}: & \,\Phi(y)\rightarrow\Phi(-y)=P\Phi(y)\,,\qquad & \nn  \\
{\cal P}': & \,\Phi(y)\rightarrow \Phi(-y+2{\pi R})=P'\Phi(y)\,, &
\eea
where $P,\,P' \equiv P T =\pm$ are \textit{orbifold parities} acting
on the field $\Phi$ in the appropriate group
representation.\footnote{Where it is assumed that $[P, T] = 0$.} The
four combinations of orbifold parities give four types of states,
with wavefunctions \bea
\zeta_m(++) \sim\cos(my/R), \nn \\
\zeta_m(+-) \sim\cos[(2m+1)y/2R], \nn \\
\zeta_m(-+) \sim\sin[(2m+1)y/2R], \nn \\
\zeta_m(--) \sim\sin[(m+1)y/R], \eea where $m\in\mathbb Z$. The
corresponding KK towers have masses
\bea M_{\rm KK}=\left\{
\begin{array}{ll}
m/R  &\,\,{\rm for}\,(PP')=(++)\,,\\
(2m+1)/2R &\,\,{\rm for}\,(PP')=(+-)\,\,\,\\
(2m+1)/2R &\,\,{\rm for}\,(PP')=(-+)\,\,\\
(m+1)/R &\,\, {\rm
for}\,(PP')=(--)\,.\label{kkmass}
\end{array}\right.
\eea
Note that only the $\Phi_{++}$ field possesses a massless zero mode.

For example, consider the Wilson line $T =  \exp (i \pi T^3) = {\rm
diag} (-1, -1, 1)$. Let $A_\mu(y) \; (A_5(y))$ have parities $P = +
(-)$, respectively.  Then only $A^3_\mu$ has orbifold parity $(++)$
and $A^3_5$ has orbifold parity $(--)$.\footnote{Note, $A^3_5(-y) =
- A^3_5(y) + \frac{1}{R}$.} Define the fields \be W^\pm =
\frac{1}{\sqrt{2}} ( A^1 \mp i A^2) \ee with $T^\pm =
\frac{1}{\sqrt{2}} ( T^1 \pm i T^2)$ and $[T^3, T^\pm] = \pm T^\pm$.
Then $W^\pm_\mu \; [W^\pm_5]$ have orbifold parity $(+ -) \; [(-
+)]$, respectively. Thus the $SO(3)$ gauge group is broken to $SO(2)
\approx U(1)$ in 4d.  The local gauge parameters preserve the $(P,
T)$ parity/holonomy, i.e. \bea
\theta^3(x_\mu, y) = \theta^3_m(x_\mu) \zeta_m(++) \nn \\
\theta^{1,2}(x_\mu, y) = \theta^{1,2}_m(x_\mu) \zeta_m(+-). \eea
 Therefore $SO(3)$ is {\em not} the symmetry at $y = \pi R$.

\subsection{A Supersymmetric $SU(5)$ orbifold GUT}\label{sec:so10}

Consider the 5d orbifold GUT model of
ref.~\cite{Hall:2001pg}.\footnote{For additional references on
orbifold SUSY GUTs see Ref. \cite{kawamura}.} The model has an
$SU(5)$ symmetry broken by orbifold parities to the SM gauge group
in 4d. The compactification scale $M_c = R^{-1}$ is assumed to be
much less than the cutoff scale.

The gauge field is a 5d vector multiplet ${\cal
V}=(A_M,\lambda,\lambda',\sigma)$, where $A_M,\,\sigma$ (and their
fermionic partners $\lambda, \ \lambda'$) are in the adjoint
representation (${\bf 24}$) of $SU(5)$ . This multiplet consists of
one 4d ${\mathcal N}= 1$ supersymmetric vector multiplet
$V=(A_\mu,\lambda)$ and one 4d chiral multiplet
$\Sigma=((\sigma+iA_5)/\sqrt{2},\lambda')$. We also add two 5d
hypermultiplets containing the Higgs doublets, ${\cal H}=(H_5,
{H_5}^c)$, $\overline{\cal H}=(\bar H_{\bar 5}, {\bar H_{\bar
5}}^c)$. The 5d gravitino $\Psi_M=(\psi^1_M,\psi_M^2)$ decomposes
into two 4d gravitini $\psi_\mu^1$, $\psi_\mu^2$ and two dilatini
$\psi_5^1$, $\psi_5^2$. To be consistent with the 5d supersymmetry
transformations one can assign positive parities to
$\psi_\mu^1+\psi_\mu^2$, $\psi_5^1-\psi_5^2$ and negative parities
to $\psi_\mu^1-\psi_\mu^2$, $\psi_5^1+\psi_5^2$; this assignment
partially breaks ${\mathcal N}=2$ to ${\mathcal N}=1$ in 4d.

The orbifold parities for various states in the vector and hyper
multiplets are chosen as follows \cite{Hall:2001pg} (where we have
decomposed all the fields into SM irreducible representations and
under $SU(5)$ we have taken \\ $P = (+++++),$  $P^\prime = (---++)$)
\bea \setlength{\arraycolsep}{0.20in}
\begin{array}{llll}
\hline
{\rm States} & P  P' &{\rm States} & P  P'\\
 \hline
V({\bf 8,1,0}) & +  + & \Sigma({\bf 8,1,0}) & -  -\\
V({\bf 1,3,0})  & +  + & \Sigma({\bf 1,3,0}) & -  -\\
V({\bf 1,1,0})  & +  + & \Sigma({\bf 1,1,0}) & -  -\\
V({\bf \bar 3,2,5/3})  & +  - & \Sigma({\bf 3,2,-5/3}) & -  + \\
V({\bf 3,2,-5/3})  & +  - & \Sigma({\bf \bar 3,2,5/3}) & -  + \\
T({\bf 3,1,-2/3})  & +  - & T^c({\bf \bar 3,1,2/3}) & -  + \\
H({\bf 1,2,+1})  & +  + & H^c({\bf \bar 1,2,-1}) & -  - \\
\bar T({\bf \bar 3,1,+2/3})  & +  - & \bar T^c({\bf 3,1,-2/3}) & -  + \\
\bar H({\bf 1,2,-1})  & +  + & \bar H^c({\bf 1,2,+1}) & -  - \\
\hline
\end{array}\,.
\eea
We see the fields supported at the orbifold fixed points $y=0$ and
$\pi R$ have parities $P=+$ and $P'=+$ respectively. They form
complete representations under the $SU(5)$ and SM groups; the
corresponding fixed points are called $SU(5)$ and SM ``branes.'' In
a 4d effective theory one would integrate out all the massive
states, leaving only massless modes of the $P=P'=+$ states. With the
above choices of orbifold parities, the SM gauge fields and the $H$
and $\bar H$ chiral multiplet are the only surviving states in 4d.
We thus have an ${\mathcal N} = 1$ SUSY in 4d.   In addition, the $T
+ \bar T$ and $T^c + \bar T^c$ color-triplet states are projected
out, solving the doublet-triplet splitting problem that plagues
conventional 4d GUTs.

\subsection{Gauge Coupling Unification}

We follow the field theoretical analysis in
ref.~\cite{Dienes:1998vg} (see also
\cite{Contino:2001si,Ghilencea:2002ff}). It has been shown there the
correction to a generic gauge coupling due to a tower of KK states
with masses $M_{\rm KK}=m/R$ is
\be
\alpha^{-1}(\Lambda)=\alpha^{-1}(\mu_0)+ \frac{b}{4\pi}
\int_{r\Lambda^{-2}}^{r\mu_0^{-2}}\frac{{\rm d}t}{t}\,
\theta_3\left(\frac{{\rm i}t}{\pi R^2}\right)\,,
\ee
where the integration is over the Schwinger parameter $t$, $\mu_0$
and $\Lambda$ are the IR and UV cut-offs, and $r=\pi/4$ is a
numerical factor. $\theta_3$ is the Jacobi theta function,
$\theta_3(t)=\sum_{m=-\infty}^\infty {\rm e}^{{\rm i}\pi m^2 t}$,
representing the summation over KK states.

For our $S^1/\mathbb{Z}_2$ orbifold there is one modification in the
calculation.   There are four sets of KK towers, with mass $M_{\rm
KK}=m/R$ (for $P=P'=+$), $(m+1)/R$ (for $P=P'=-$) and $(m+1/2)/R$
(for $P=+$, $P'=-$ and $P=-$, $P'=+$), where $m\geq 0$. The
summations over KK states give respectively
$\frac{1}{2}\left(\theta_3({\rm i}t/\pi R^2)-1\right)$ for the first
two cases and $\frac{1}{2}\theta_2({\rm i}t/\pi R^2)$ for the last
two (where $\theta_2(t)= \sum_{m=-\infty}^\infty {\rm e}^{{\rm i}\pi
(m+1/2)^2 t}$), and we have separated out the zero modes in the
$P=P'=+$ case.

Tracing the renormalization group evolution from low energy scales,
we are first in the realm of the MSSM, and the beta function
coefficients are ${\bf b}^{MSSM}=(- \frac{33}{5},-1,3)$. The next
energy threshold is the compactification scale $M_c$. From this
scale to the cut-off scale, $M_*$, we have the four sets of KK
states.

Collecting these facts, and using $\theta_2({\rm i}t/\pi R^2)\simeq
\theta_3({\rm i}t/\pi R^2)\simeq\sqrt{\frac{\pi}{t}} R$ for
$t/R^2\ll 1$, we find the RG equations,
\bea \label{running_fiveD}    \alpha_i^{-1}(M_Z) = & & \\
& \alpha_{*}^{-1} - \frac{b_i^{MSSM}}{2\pi} \log \frac{M_*}{M_Z}  +
\frac{1}{4\pi} \left(b_i^{++} + b_i^{--}\right) \log \frac{M_*}{M_c}
& \nn \\
& - \frac{b^{\mathcal G}}{2 \pi}\left(\frac{M_*}{M_c} - 1\right) +
\delta_i^2 + \delta_i^l & \nn \eea
for $i=1,2,3$, where $\alpha_*^{-1} = \frac{8 \pi^2 R}{g_5^2}$ and
we have taken the cut-off scales, $\mu_0=M_c = \frac{1}{R}$ and
$\Lambda=M_*$.  (Note, this 5d orbifold GUT is a non-renormalizable
theory with a cut-off.   In string theory, the cut-off will be
replaced by the physical string scale, $M_{\rm STRING}$.)
$b^{\mathcal G} = \sum_{P=\pm, P'=\pm} b^{\mathcal G}_{PP'}$, so in
fact it is the beta function coefficient of the orbifold GUT gauge
group, ${\mathcal G} = SU(5)$. The beta function coefficients in the
last two terms have an ${\mathcal N}=2$ nature, since the massive KK
states enjoy a larger supersymmetry.  In general we have
$b^{\mathcal G} = 2 C_2(\mathcal G) - 2 N_{hyper} T_R$.  The first
term (in Eqn. \ref{running_fiveD}) on the right is the 5d gauge
coupling defined at the cut-off scale, the second term accounts for
the one loop RG running in the MSSM from the weak scale to the
cut-off,  the third and fourth terms take into account the KK modes
in loops above the compactification scale and the last two terms
account for the corrections due to two loop RG running and weak
scale threshold corrections.

It should be clear that there is a simple correspondence to the 4d
analysis. We have \bea \alpha_G^{-1} \;\; (4d) & \leftrightarrow
\alpha_*^{-1} - \frac{b^{\mathcal G}}{2 \pi}\left( \frac{M_*}{M_c} -
1\right) \;\; (5d) & \\ \label{eq:deltash} \delta_i^h \;\; (4d)
\leftrightarrow & \frac{1}{4\pi} \left(b_i^{++} + b_i^{--}\right)
\log \frac{M_*}{M_c} - \frac{b^{MSSM}_i}{2 \pi} \log \frac{M_*}{M_G}
\;\; (5d) .& \nn \eea Thus in 5d the GUT scale threshold corrections
determine the ratio $M_*/M_c$ (note the second term in Eqn.
\ref{eq:deltash} does not contribute to $\delta_s^h$). For $SU(5)$
we have ${\bf b^{++} + b^{--}} = ( - 6/5, 2, 6)$ and given
$\delta_s^h$ (Eqn. \ref{eq:deltah}) we have \be \delta_s^h =
\frac{12}{28 \pi} \log \frac{M_*}{M_c} \approx + 0.94 \ee or \be
\frac{M_*}{M_c} \approx 10^3. \ee

\begin{figure}
\includegraphics[height=.23\textheight]{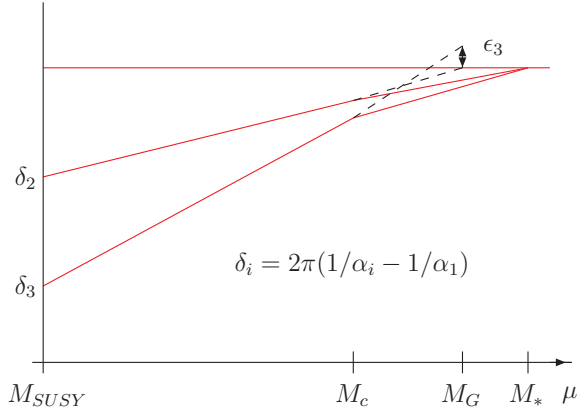}
\caption{\label{fig:gcu} The differences $\delta_i = 2 \pi(
1/\alpha_i - 1/\alpha_1)$ are plotted as a function of energy scale
$\mu$.  The threshold correction $\epsilon_3$ defined in the 4d GUT
scale is used to fix the threshold correction in the 5d orbifold
GUT. }
\end{figure}

If the GUT scale is defined at the point where $\alpha_1 =
\alpha_2$, then we have $\delta_1^h = \delta_2^h$ or $\log
\frac{M_*}{M_G} \approx 2$.  In 5d orbifold GUTs, nothing in
particular happens at the 4d GUT scale.   However, since the gauge
bosons affecting the dimension 6 operators for proton decay obtain
their mass at the compactification scale,  it is important to
realize that the compactification scale is typically lower than the
4d GUT scale and the cut-off is higher (see Figure \ref{fig:gcu}).

\subsection{Quarks and Leptons in 5d Orbifold GUTs}

Quarks and lepton fields can be put on either of the orbifold
``branes" or in the 5d bulk.   If they are placed on the $SU(5)$
``brane" at $y = 0$, then they come in complete $SU(5)$ multiplets.
As a consequence a coupling of the type \be W \supset \int d^2\theta
\int  dy \  \delta(y) \ \bar H \ 10 \ \bar 5 \ee will lead to bottom
- tau Yukawa unification.   This relation is good for the third
generation and so it suggests that the third family should reside on
the $SU(5)$ brane.   Since this relation does not work for the first
two families, they might be placed in the bulk or on the SM brane at
$y = \pi \ R$.  Without further discussion of quark and lepton
masses (see
\cite{Hall:2002ci,Kim:2004vk,Alciati:2005ur,Alciati:2006sw} for
complete $SU(5)$ or $SO(10)$ orbifold GUT models), let us consider
proton decay in orbifold GUTs.

\subsection{Proton Decay}

\subsubsection{Dimension 6 Operators}

The interactions contributing to proton decay are those between the
so-called $X$ gauge bosons $A_\mu^{(+-)} \in V(+-)$ (where
${A_{\mu}^{(+-)}}^{a i}(x_\mu,y)$ is the five dimensional gauge
boson with quantum numbers $(\bar 3,2,+5/3)$ under SU(3)$\times$
SU(2) $\times$ U(1), $a$ and $i$ are color and SU(2) indices
respectively) and the ${\mathcal N}=1$ chiral multiplets on the
$SU(5)$ brane at $y=0$. Assuming all quarks and leptons reside on
this brane we obtain the $\Delta B \neq 0$ interactions given by \be
{\cal S}_{\Delta B \neq 0} = - \frac{g_5}{\sqrt{2}} \int d^4x
{A_{\mu}^{(+-)}}^{a i}(x_\mu,0) J_{a i}^{\mu}(x_\mu)+ h.c.~~~. \ee
The currents $J_{a i}^{\mu}$ are given by: \bea J_{a i}^{\mu}&=&
\epsilon_{a b c}\,\epsilon_{i j} (u^{c})^*_b \,\bar \sigma^\mu \,
q^{c j} +  q_{a i}^* \, \bar \sigma^\mu\, e^c - \tilde l_{i}^* \, \bar \sigma^\mu\, (d^c)_a \nn\\
&=&  (u^c)^* \, \bar \sigma^\mu\, q + q^* \, \bar \sigma^\mu\, e^c -
\tilde l^* \, \bar \sigma^\mu\, d^c  ~~~, \eea

Upon integrating out the $X$ gauge bosons we obtain the effective
lagrangian for proton decay  \be {\cal L} = -\frac{g_G^2}{2 M_X^2}
\sum_{i, j} \left[({q^*_i} \bar \sigma^\mu u^c_i) \; (\tilde l^*_j
\bar \sigma_\mu d^c_j) \ + \ (q^*_i \bar \sigma^\mu e^c_i) \; (q^*_j
\bar \sigma_\mu u^c_j) \right]~~~, \label{lagpd0} \ee where all
fermions are weak interaction eigenstates and $i,j,k=1,2,3$ are
family indices. The dimensionless quantity \be g_G\equiv g_5
\frac{1}{\sqrt{2 \pi R}} \ee is the four-dimensional gauge coupling
of the gauge bosons zero modes. The combination \be
M_X=\frac{M_c}{\pi} ~~~, \ee proportional to the compactification
scale \be M_c\equiv \frac{1}{R}~~~, \ee is an effective gauge vector
boson mass arising from the sum over all the Kaluza-Klein levels:
\be \sum_{n=0}^\infty\frac{4}{(2 n+1)^2 M_c^2}=\frac{1}{M_X^2}~~~.
\ee Before one can evaluate the proton decay rate one must first
rotate the quark and lepton fields to a mass eigenstate basis.  This
will bring in both left- and right-handed quark and lepton mixing
angles. However, since the compactification scale is typically lower
than the 4d GUT scale, it is clear that proton decay via dimension 6
operators is likely to be enhanced.

\subsubsection{Dimension 5 Operators}

The dimension 5 operators for proton decay result from integrating
out color triplet Higgs fermions.   However in this simplest $SU(5)$
5d model the color triplet mass is of the form
\cite{ArkaniHamed:2001tb} \be W \supset \int d^2 \theta \ dy \
(T(-+)^c \
\partial_y \ T(+-) + \bar T(-+)^c \ \partial_y \ \bar T(+-)) \ee where a sum over massive
KK modes is understood. Since only $T, \ \bar T$ couple directly to
quarks and leptons,  {\em no dimension 5 operators are obtained when
integrating out the color triplet Higgs fermions}.

\subsubsection{Dimension 4 baryon and lepton violating operators}

If the theory is constructed with an R parity or family reflection
symmetry, then no such operators will be generated.

\section{Heterotic String Orbifolds and Orbifold GUTs}

\section{Phenomenological guidelines}
We use the following guidelines when searching for ``realistic"
string models~\cite{Lebedev:2006kn,Lebedev:2007hv}. We want to:
\begin{enumerate}
\item Preserve gauge coupling unification; \item  Low energy SUSY as solution to the gauge hierarchy problem,
i.e. why is $M_Z << M_G$; \item Put quarks and leptons in {\bf 16}
of SO(10); \item Put Higgs in {\bf 10}, thus quarks and leptons are
distinguished from Higgs by their SO(10) quantum numbers; \item
Preserve GUT relations for 3rd family Yukawa couplings; \item  Use
the fact that GUTs accommodate a ``Natural" See-Saw scale ${\cal
O}(M_G)$;  \item Use intuition derived from Orbifold GUT
constructions, \cite{Kobayashi:2004ud,Kobayashi:2004ya} and \item
Use local GUTs to enforce family structure
\cite{Forste:2004ie,Buchmuller:2005jr,Buchmuller:2006ik}.
\end{enumerate}
It is the last two guidelines which are novel and characterize our
approach.  As a final comment, the string theory analysis discussed
here assumes supersymmetric vacua at the string scale.   As a
consequence there are generically a multitude of moduli.  The gauge
and Yukawa couplings depend on the values of the moduli vacuum
expectation values [VEVs].  In addition vector-like
exotics\footnote{By definition, a vector-like exotic can obtain mass
without breaking any Standard Model gauge symmetry.} can have mass
proportional to the moduli VEVs.  We will assume arbitrary values
for these moduli VEVs along supersymmetric directions, in order to
obtain desirable low energy phenomenology.   Of course, at some
point supersymmetry must be broken and these moduli must be
stabilized.   We save this harder problem for a later date.
Nevertheless, we can add one more guideline at this point.   In the
supersymmetric limit,  we want the superpotential to have a
vanishing VEV.   This is so that we can work in flat Minkowski space
when considering supergravity.   Some of our models naturally have
this property.

\subsection{$E_8 \times E_8$ 10D heterotic string compactified on $\mz_3 \times \mz_2$ 6D orbifold}

There are many reviews and books on string theory.  I cannot go into
great detail here, so I will confine my discussion to some basic
points.  We start with the 10d heterotic string theory, consisting
of a 26d left-moving bosonic string and a 10d right-moving
superstring. Modular invariance requires the momenta of the internal
left-moving bosonic degrees of freedom (16 of them) to lie in a 16d
Euclidean even self-dual lattice, we choose to be the
$\lgp{E}{8}\times\lgp{E}{8}$ root lattice.\footnote{For an
orthonormal basis, the $\lgp{E}{8}$ root lattice consists of
following vectors, $(n_1,n_2,\cdots,n_8)$ and
$(n_1+\frac{1}{2},n_2+\frac{1}{2},\cdots,n_8+\frac{1}{2})$, where
$n_1, n_2,\cdots n_8$ are integers and $\sum_{i=1}^8 n_i=0\,{\rm
mod}\,2$. }

\subsubsection{Heterotic string compactified on $T^6/\mz_6$}

\begin{figure*}
\scalebox{0.85}{\includegraphics{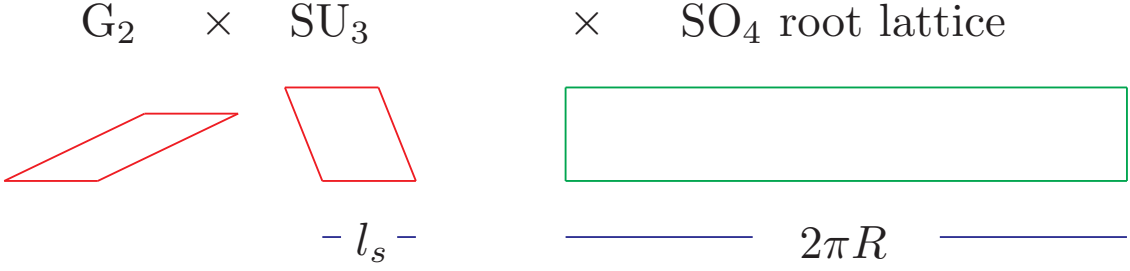}} \caption{$G_2 \oplus SU(3)
\oplus SO(4)$ lattice. Note, we have taken 5 directions with string
scale length $\ell_s$ and one with length $2 \pi R \gg \ell_s$. This
will enable the analogy of an effective 5d orbifold field theory.
\label{fig:untwisted} }
\end{figure*}

We first compactify the theory on 6d torus defined by the space
group action of translations on a factorizable Lie algebra lattice
$G_2\oplus SU(3)\oplus SO(4)$ (see Fig. \ref{fig:untwisted}). Then
we mod out by the $\mz_6$ action on the three complex compactified
coordinates given by $Z^i\rightarrow e^{2\pi i {\bf r}_i\cdot{\bf
v}_6}Z^i$, $i=1,2,3$, where ${\bf v}_6=\frac{1}{6}(1,2,-3)$ is the
twist vector, and ${\bf r}_1=(1,0,0,0)$, ${\bf r}_2=(0,1,0,0)$,
${\bf r}_3=(0,0,1,0)$.\footnote{Together with ${\bf r}_4=(0,0,0,1)$,
they form the set of positive weights of the ${\bf 8}_v$
representation of the $SO(8)$, the little group in 10d. $\pm{\bf
r}_4$ represent the two uncompactified dimensions in the light-cone
gauge. Their space-time fermionic partners have weights ${\bf
r}=(\pm\frac{1}{2},\pm\frac{1}{2},\pm\frac{1}{2},\pm\frac{1}{2})$
with even numbers of positive signs; they are in the ${\bf 8}_s$
representation of $SO(8)$. In this notation, the fourth component of
${\bf v}_6$ is zero.\label{fn1}} For simplicity and definiteness, we
also take the compactified space to be a factorizable Lie algebra
lattice $G_2\oplus SU(3)\oplus SO(4)$ (see Fig.
\ref{fig:untwisted}).

The $\mz_6$ orbifold is equivalent to a $\mz_2\times\mz_3$ orbifold,
where the two twist vectors are ${\bf v}_2=3{\bf
v}_6=\frac{1}{2}(1,0,-1)$ and ${\bf v}_3=2{\bf
v}_6=\frac{1}{3}(1,-1,0)$. The $\mz_2$ and $\mz_3$ sub-orbifold
twists have the $SU(3)$ and $SO(4)$ planes as their fixed torii. In
Abelian symmetric orbifolds, gauge embeddings of the point group
elements and lattice translations are realized by shifts of the
momentum vectors, ${\bf P}$, in the $E_8\times E_8$ root
lattice\footnote{The $E_8$ root lattice is given by the set of
states ${\bf P} = \{n_1, n_2, \cdots, n_8 \}, \ \{n_1 +\frac{1}{2},
n_2 +\frac{1}{2}, \cdots, n_8 +\frac{1}{2} \}$ satisfying $n_i \in
\mz, \ \sum_{i = 1}^8 n_i = 2 \mz$.}~\cite{IMNQ}, {\it i.e.}, ${\bf
P}\rightarrow {\bf P}+k{\bf V}+l{\bf W}$, where $k, l$ are some
integers, and ${\bf V}$ and ${\bf W}$ are known as the gauge twists
and Wilson lines \cite{wl}. These embeddings are subject to modular
invariance requirements \cite{Dixon,vafa}. The Wilson lines are also
required to be consistent with the action of the point group. In the
$\mz_6$ model, there are at most three consistent Wilson lines
\cite{kobayashi}, one of degree 3 (${\bf W}_3$), along the $SU(3)$
lattice, and two of degree 2 (${\bf W}_2,\,{\bf W}_2'$), along the
$SO(4)$ lattice.

\begin{figure*}
\scalebox{0.85}{\includegraphics{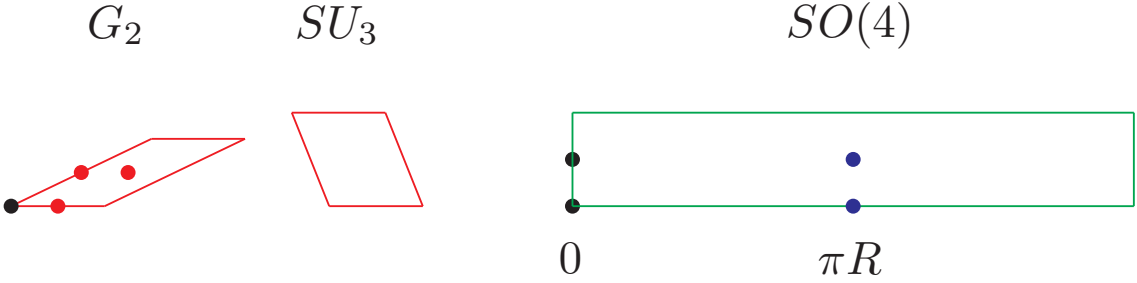}} \caption{$G_2 \oplus SU(3)
\oplus SO(4)$ lattice with $\mz_2$ fixed points. The $T_{3}$ twisted
sector states sit at these fixed points.  The fixed point at the
origin and the symmetric linear combination of the red (grey) fixed
points in the $G_2$ torus have $\gamma =1$. \label{fig:T3} }
\end{figure*}

\begin{figure*}
\scalebox{0.85}{
\includegraphics{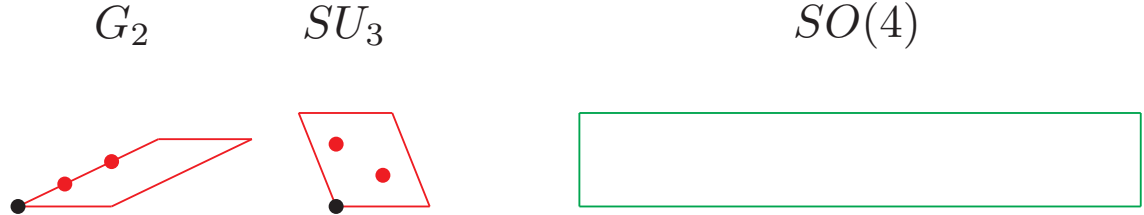}}
\caption{\label{fig:T2} $G_2 \oplus SU(3) \oplus SO(4)$ lattice with
$\mz_3$ fixed points for the $T_2$ twisted sector.  The fixed point
at the origin and the symmetric linear combination of the red (grey)
fixed points in the $G_2$ torus have $\gamma =1$.}
\end{figure*}

\begin{figure*}
\scalebox{0.85}{
\includegraphics{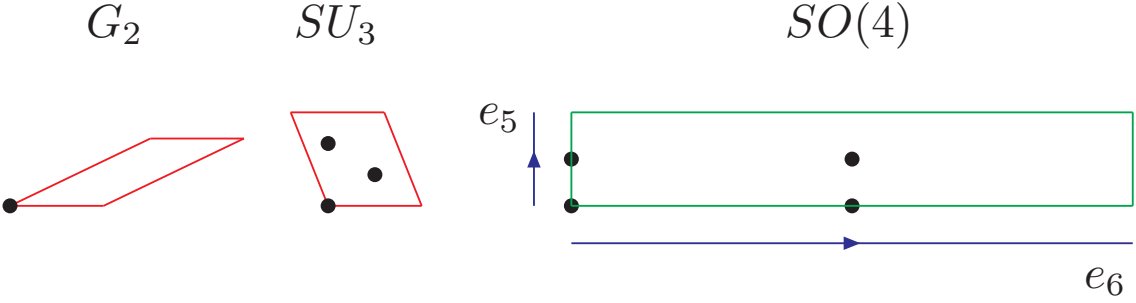}}
\caption{\label{fig:T1} $G_2 \oplus SU(3) \oplus SO(4)$ lattice with
$\mz_6$ fixed points. The $T_{1}$ twisted sector states sit at these
fixed points.}
\end{figure*}

The $\mz_6$ model has three untwisted sectors ($U_i,\,i=1,2,3$) and
five twisted sectors ($T_i,\,i=1,2,\cdots,5$). (The $T_k$ and
$T_{6-k}$ sectors are CPT conjugates of each other.) The twisted
sectors split further into sub-sectors when discrete Wilson lines
are present. In the $SU(3)$ and $SO(4)$ directions, we can label
these sub-sectors by their winding numbers, $n_3=0,1,2$ and
$n_2,\,n'_2=0,1$, respectively. In the $G_2$ direction, where both
the $\mz_2$ and $\mz_3$ sub-orbifold twists act, the situation is
more complicated.  There are four $\mz_2$ fixed points in the $G_2$
plane. Not all of them are invariant under the $\mz_3$ twist, in
fact three of them are transformed into each other. Thus for the
$T_3$ twisted-sector states one needs to find linear combinations of
these fixed-point states such that they have definite eigenvalues,
$\gamma=1$ (with multiplicity 2), $e^{i2\pi/3}$, or $e^{i4\pi/3}$,
under the orbifold twist \cite{DFMS,kobayashi} (see
Fig.~\ref{fig:T3}). Similarly, for the $T_{2,4}$ twisted-sector
states, $\gamma=1$ (with multiplicity 2) and $-1$ (the fixed points
of the $T_{2,4}$ twisted sectors in the $G_2$ torus are shown in
Fig.~\ref{fig:T2}). The $T_{1}$ twisted-sector states have only one
fixed point in the $G_2$ plane, thus $\gamma=1$ (see
Fig.~\ref{fig:T1}). The eigenvalues $\gamma$ provide another piece
of information to differentiate twisted sub-sectors.


Massless states in 4d string models consist of those momentum
vectors ${\bf P}$ and ${\bf r}$ ($\bf r$ are in the $SO(8)$ weight
lattice) which satisfy the following mass-shell equations
\cite{Dixon,IMNQ}, %%
\bea %%
&&\frac{\alpha'}{2}m_{R}^2=N^k_R+\frac{1}{2}\left|{\bf r}+k{\bf v}\right|^2+a^k_R=0\,,\label{masscond1}\\
&&\frac{\alpha'}{2}m_L^2=N_L^k+\frac{1}{2}\left|{\bf P}+k{\bf
X}\right|^2+a_L^k=0\,, \label{masscond2} %%
\eea %%
where $\alpha'$ is the Regge slope, $N^k_R$ and $N^k_L$ are
(fractional) numbers of the right- and left-moving (bosonic)
oscillators, ${\bf X}={\bf V}+n_3{\bf W}_3+n_2{\bf W}_2+n_2'{\bf
W}_2'$, and $a^k_R$, $a_L^k$ are the normal ordering constants, %%
\bea %%
a^k_R
&=&-\frac{1}{2}+\frac{1}{2}\sum_{i=1}^3|{\widehat{kv_i}}|\left(1-|{\widehat{kv_i}}|\right)\,,\nn\\
a_L^k
&=&-1+\frac{1}{2}\sum_{i=1}^3|{\widehat{kv_i}}|\left(1-|{\widehat{kv_i}}|\right)\,,
\eea %%
with $\widehat{kv_i}={\rm mod}(kv_i, 1)$.

These states are subject to a generalized Gliozzi-Scherk-Olive (GSO)
projection ${\cal
P}=\frac{1}{6}\sum_{\ell=0}^{5}\Delta^\ell$~\cite{IMNQ}. For the
simple case of the $k$-th twisted sector ($k=0$ for the untwisted
sectors) with no Wilson lines ($n_3 = n_2 = n^\prime_2 = 0$) we have
\be
\Delta= \gamma\phi  \exp\left\{i\pi \biggl[(2{\bf P}+k{\bf X})
\cdot{\bf X} -(2{\bf r}+k{\bf v})\cdot {\bf
v}\biggr]\right\},\label{GSO}
\ee
where $\phi$ are phases from bosonic oscillators.   However, in the
$\mz_6$ model, the GSO projector must be modified for the
untwisted-sector and $T_{2,4}$, $T_3$ twisted-sector states in the
presence of Wilson lines \cite{Kobayashi:2004ya}. The Wilson lines
split each twisted sector into sub-sectors and there must be
additional projections with respect to these sub-sectors. This
modification in the projector gives the following projection
conditions,
\be
{\bf P}\cdot{\bf V}-{\bf r}_i\cdot{\bf v}=\mz\,\,\,\,(i=1,2,3),\quad
{\bf P}\cdot{\bf W}_3,\,\,{\bf P}\cdot{\bf W}_2,\,\,{\bf P}\cdot{\bf
W}_2'=\mz, \label{eq1}
\ee
for the untwisted-sector states, and
\be
T_{2,4}:\,{\bf P}\cdot{\bf W}_2,\,\,{\bf P}\cdot{\bf W}_2'={\mathbb
Z}\,,\qquad T_3:\,{\bf P}\cdot{\bf W}_3={\mathbb Z}\,,\label{eq2}
\ee
for the $T_{2,3,4}$ sector states (since twists of these sectors
have fixed torii). There is no additional condition for the $T_1$
sector states.

\subsubsection{An orbifold GUT -- heterotic string dictionary}

We first implement the $\mz_3$ sub-orbifold twist, which acts only
on the $G_2$ and $SU(3)$ lattices. The resulting model is a 6d gauge
theory with ${\cal N}=2$ hypermultiplet matter, from the untwisted
and $T_{2,4}$ twisted sectors. This 6d theory is our starting point
to reproduce the orbifold GUT models.  The next step is to implement
the $\mz_2$ sub-orbifold twist. The geometry of the extra dimensions
closely resembles that of 6d orbifold GUTs. The $SO(4)$ lattice has
four $\mz_2$ fixed points at $0$, $\pi R$, $\pi R^\prime$ and $\pi(R
+ R^\prime)$, where $R$ and $R'$ are on the $e_5$ and $e_6$ axes,
respectively, of the lattice (see Figs.~\ref{fig:T3} and
\ref{fig:T1}). When one varies the modulus parameter of the $SO(4)$
lattice such that the length of one axis ($R$) is much larger than
the other ($R'$) and the string length scale ($\ell_s$), the lattice
effectively becomes the $S^1/\mz_2$ orbi-circle in the 5d orbifold
GUT, and the two fixed points at $0$ and $\pi R$ have degree-2
degeneracies. Furthermore, one may identify the states in the
intermediate $\mz_3$ model, {\it i.e.} those of the untwisted and
$T_{2,4}$ twisted sectors, as bulk states in the orbifold GUT.

Space-time supersymmetry and GUT breaking in string models work
exactly as in the orbifold GUT models.  First consider supersymmetry
breaking. In the field theory, there are two gravitini in 4d, coming
from the 5d (or 6d) gravitino. Only one linear combination is
consistent with the space reversal, $y\rightarrow -y$; this breaks
the ${\cal N}=2$ supersymmetry to that of ${\cal N}=1$. In string
theory, the space-time supersymmetry currents are represented by
those half-integral $SO(8)$ momenta.\footnote{Together with ${\bf
r}_4=(0,0,0,1)$, they form the set of positive weights of the ${\bf
8}_v$ representation of the $SO(8)$, the little group in 10d.
$\pm{\bf r}_4$ represent the two uncompactified dimensions in the
light-cone gauge. Their space-time fermionic partners have weights
${\bf
r}=(\pm\frac{1}{2},\pm\frac{1}{2},\pm\frac{1}{2},\pm\frac{1}{2})$
with even numbers of positive signs; they are in the ${\bf 8}_s$
representation of $SO(8)$. In this notation, the fourth component of
${\bf v}_6$ is zero.} The $\mz_3$ and $\mz_2$ projections remove all
but two of them, ${\bf r}=\pm\frac{1}{2}(1,1,1,1)$; this gives
${\cal N}=1$ supersymmetry in 4d.

Now consider GUT symmetry breaking. As usual, the $\mz_2$ orbifold
twist and the translational symmetry of the $SO(4)$ lattice are
realized in the gauge degrees of freedom by degree-2 gauge twists
and Wilson lines respectively. To mimic the 5d orbifold GUT example,
we impose only one degree-2 Wilson line, ${\bf W}_2$, along the long
direction of the $SO(4)$ lattice, ${\bf R}$.\footnote{Wilson lines
can be used to reduce the number of chiral families. In all our
models, we find it is sufficient to get three-generation models with
two Wilson lines, one of degree 2 and one of degree 3. Note,
however, that with two Wilson lines in the $SO(4)$ torus we can
break $SO(10)$ directly to $SU(3) \times SU(2) \times U(1)_Y \times
U(1)_X$ (see for example, Ref.~\cite{6dOGUT}).} The gauge embeddings
generally break the 5d/6d (bulk) gauge group further down to its
subgroups, and the symmetry breaking works exactly as in the
orbifold GUT models. This can clearly be seen from the following
string theoretical realizations of the orbifold parities
\be P=p\,e^{2\pi i\,[{\bf P}\cdot{\bf V}_2-{\bf r}\cdot{\bf
v}_2]}\,,\quad P'=p\,e^{2\pi i\,[{\bf P}\cdot({\bf V}_2+{\bf
W}_2)-{\bf r}\cdot{\bf v}_2]}\,,\label{PP'}
\ee
where ${\bf V}_2=3{\bf V}_6$, and $p=\gamma\phi$ can be identified
with intrinsic parities in the field theory language.\footnote{For
gauge and untwisted-sector states, $p$ are trivial. For
non-oscillator states in the $T_{2,4}$ twisted sectors, $p=\gamma$
are the eigenvalues of the $G_2$-plane fixed points under the
${\mathbb Z}_2$ twist. Note that $p=+$ and $-$ states have
multiplicities $2$ and $1$ respectively since the corresponding
numbers of fixed points in the $G_2$ plane are $2$ and $1$.} Since
$2({\bf P}\cdot{\bf V}_2-{\bf r}\cdot{\bf v}_2),\,2{\bf P}\cdot{\bf
W}_2=\mz$, by properties of the $E_8\times E_8$ and $SO(8)$
lattices, thus $P^2=P'^2=1$, and Eq.~(\ref{PP'}) provides a
representation of the orbifold parities. From the string theory
point of view, $P=P'=+$ are nothing but the projection conditions,
$\Delta=1$, for the untwisted and $T_{2,4}$ twisted-sector states
(see Eqs.~(\ref{GSO}), (\ref{eq1}) and (\ref{eq2})).

To reaffirm this identification, we compare the masses of KK
excitations derived from string theory with that of orbifold GUTs.
The coordinates of the $SO(4)$ lattice are untwisted under the
$\mz_3$ action, so their mode expansions are the same as that of
toroidal coordinates. Concentrating on the ${\bf R}$ direction, the
bosonic coordinate is
$X_{L,R}=x^{}_{L,R}+p^{}_{L,R}(\tau\pm\sigma)+{\rm oscillator\,
terms}$, with $p^{}_L$, $p^{}_R$ given by
\bea
p^{}_L= & \frac{m}{2R}+\left(1-\frac{1}{4}|{\bf W}_2|^2\right)
\frac{n_2R}{\ell_s^2}+\frac{{\bf P}\cdot{\bf W}_2}{2R} \,,\qquad \nn
\\ p^{}_R= & p^{}_L-\frac{2n_2R}{\ell_s^2}\,,\label{PLR} &
\eea
where $m$ ($n_2$) are KK levels (winding numbers). The ${\mathbb
Z}_2$ action maps $m$ to $-m$, $n_2$ to $-n_2$ and ${\bf W}_2$ to
$-{\bf W}_2$, so physical states must contain linear combinations,
$|m,n_2\rangle\pm|-m,-n_2\rangle$; the eigenvalues $\pm 1$
correspond to the first ${\mathbb Z}_2$ parity, $P$, of orbifold GUT
models. The second orbifold parity, $P'$, induces a non-trivial
degree-2 Wilson line; it shifts the KK level by $m\rightarrow m+{\bf
P}\cdot{\bf W}_2$. Since $2{\bf W}_2$ is a vector of the (integral)
$E_8\times E_8$ lattice, the shift must be an integer or
half-integer. When $R\gg R'\sim\ell_s$, the winding modes and the KK
modes in the smaller dimension of $SO(4)$ decouple.  Eq.~(\ref{PLR})
then gives four types of KK excitations, reproducing the field
theoretical mass formula in Eq.~(\ref{kkmass}).

\subsection{MSSM with R parity}

In this section we discuss just one ``benchmark" model (Model 1)
obtained via a ``mini-landscape" search \cite{Lebedev:2006kn} of the
$E_8 \times E_8$ heterotic string compactified on the $\mz_6$
orbifold \cite{Lebedev:2007hv}.\footnote{For earlier work on MSSM
models from $\mz_6$ orbifolds of the heterotic string, see
\cite{Buchmuller:2005jr,Buchmuller:2006ik}.} The model is defined by
the shifts and Wilson lines
\begin{subequations}
\begin{eqnarray}
 V & = &
 \left( \frac{1}{3},-\frac{1}{2},-\frac{1}{2},0,0,0,0,0\right)\,
   \left(\frac{1}{2},-\frac{1}{6},-\frac{1}{2},-\frac{1}{2},-\frac{1}{2},-\frac{1}{2},-\frac{1}{2},\frac{1}{2}\right)\;, \nn \\
 W_2 & = &
 \left( 0,-\frac{1}{2},-\frac{1}{2},-\frac{1}{2},\frac{1}{2},0,0,0\right)\,
   \left(
   4,-3,-\frac{7}{2},-4,-3,-\frac{7}{2},-\frac{9}{2},\frac{7}{2}\right)\;, \nn \\
 W_3 & = &
 \left(-\frac{1}{2},-\frac{1}{2},\frac{1}{6},\frac{1}{6},\frac{1}{6},\frac{1}{6},\frac{1}{6},\frac{1}{6}\right)\,
   \left( \frac{1}{3},0,0,\frac{2}{3},0,\frac{5}{3},-2,0\right)\;.
\end{eqnarray}
\end{subequations}
A possible second order 2 Wilson line is set to zero.

The shift $V$ is defined to satisfy two criteria. \begin{itemize}
\item The first criterion is the existence of a local $SO(10)$ GUT \footnote{For more discussion on local GUTs, see
\cite{Forste:2004ie,Buchmuller:2005jr}} at the $T_1$ fixed points at
$x_6 =0$ in the $SO(4)$ torus (Fig. \ref{fig:T1}). \be P \cdot V =
\mathbb{Z};  \;  P \in SO(10) \;\; {\rm momentum \; lattice} .\ee
Since the $T_1$ twisted sector has no invariant torus and only one
Wilson line along the $x_6$ direction,  all states located at these
two fixed points must come in complete $SO(10)$ multiplets.
\item  The second criterion is that two massless spinor representations of $SO(10)$ are
located at the $x_6 = 0$ fixed points.
\end{itemize}
Hence, the two complete families on the local $SO(10)$ GUT fixed
points gives us an excellent starting point to find the MSSM.  The
Higgs doublets and third family of quarks and leptons must then come
from elsewhere.

Let us now discuss the effective 5d orbifold GUT
\cite{Dundee:2008ts}.  Consider the orbifold $(T^2)^3/\mz_3$ plus
the Wilson line $W_3$ in the $SU_3$ torus.  The $\mz_3$ twist does
not act on the $SO_4$ torus, see Fig. \ref{fig:T2}. As a consequence
of embedding the $\mz_3$ twist as a shift in the $E_8 \times E_8$
group lattice and taking into account the $W_3$ Wilson line, the
first $E_8$ is broken to $SU(6)$.  This gives the effective 5d
orbifold gauge multiplet contained in the ${\mathcal N}=1$ vector
field $V$.  In addition we find the massless states $\Sigma \in {\bf
35}$, ${\bf 20} + {\bf 20}^c$ and 18 (${\bf 6} + {\bf 6}^c$) in the
6d untwisted sector and $T_2, \ T_4$ twisted sectors. Together these
form a complete ${\mathcal N}=2$ gauge multiplet ($V + \Sigma$) and
a {\bf 20} + 18 ({\bf 6}) dimensional hypermultiplets. In fact the
massless states in this sector can all be viewed as ``bulk" states
moving around in a large 5d space-time.

Now consider the $\mz_2$ twist and the Wilson line $W_2$ along the
$x_6$ axis in the $SO_4$ torus.   The action of the $\mz_2$ twist
breaks the gauge group to $SU(5)$, while $W_2$ breaks $SU(5)$
further to the SM gauge group $SU(3)_C \times SU(2)_L \times
U(1)_Y$.

Let us now consider those MSSM states located in the bulk.  From two
of the pairs of ${\mathcal N}=1$ chiral multiplets $\bf{6} +
\bf{6}^c$, which
decompose as \bea & 2 \times (\bf 6 + \bf 6^c) \supset  &  \\
& \left[(1,\bf 2)_{1,1}^{--}+(\bf 3,1)_{-2/3,1/3}^{-+}\right] & \nn
\\ & + \left[(1,\bf 2)_{-1,-1}^{++}+(\overline{\bf
3},1)_{2/3,-1/3}^{--}\right] & \nonumber
\\
& +\left[(1,\bf 2)_{1,1}^{-+}+(\bf 3,1)_{-2/3,1/3}^{--}\right] & \nn
\\ & + \left[(1,\bf 2)_{-1,-1}^{+-}+(\overline{\bf
3},1)_{2/3,-1/3}^{++}\right], & \nonumber  \eea we obtain the third
family $b^c$ and lepton doublet, $l$. The rest of the third family
comes from the $\mathbf{10} + \mathbf{10}^c$ of $SU(5)$ contained in
the $\mathbf{20} + \mathbf{20}^c$ of $SU(6)$, in the untwisted
sector.

Now consider the Higgs bosons.  The bulk gauge symmetry is $SU(6)$.
Under $SU(5) \times U(1)$, the adjoint decomposes as
\begin{equation}
   \mathbf{35} \rightarrow \mathbf{24}_0 + \mathbf{5}_{+1} + \mathbf{5}^c_{-1} + 1_0.
\end{equation}
Thus the MSSM Higgs sector emerges from the breaking of the $SU(6)$
adjoint by the orbifold and the model satisfies the property of
``gauge-Higgs unification."

In the models with gauge-Higgs unification, the Higgs multiplets
come from the 5d \textit{vector} multiplet ($V, \Sigma$), both in
the adjoint representation of $SU(6)$.   $V$ is the 4d gauge
multiplet and the 4d chiral multiplet $\Sigma$ contains the Higgs
doublets. These states transform as follows under the orbifold
parities $(P\,\,P')$:
\begin{equation}
  V:\: \left( \begin{array}{ccc|cc|c}
    (+ +) & (+ +) & (+ +) & (+ -) & (+ -) & (- +) \\
    (+ +) & (+ +) & (+ +) & (+ -) & (+ -) & (- +) \\
    (+ +) & (+ +) & (+ +) & (+ -) & (+ -) & (- +) \\ \hline
    (+ -) & (+ -) & (+ -) & (+ +) & (+ +) & (- -) \\
    (+ -) & (+ -) & (+ -) & (+ +) & (+ +) & (- -) \\ \hline
    (- +) & (- +) & (- +) & (- -) & (- -) & (+ +)
  \end{array} \right)
\label{eq:V6trans}
\end{equation}
\begin{equation}
  \Phi:\: \left( \begin{array}{ccc|cc|c}
    (- -) & (- -) & (- -) & (- +) & (- +) & (+ -) \\
    (- -) & (- -) & (- -) & (- +) & (- +) & (+ -) \\
    (- -) & (- -) & (- -) & (- +) & (- +) & (+ -) \\ \hline
    (- +) & (- +) & (- +) & (- -) & (- -) & (+ +) \\
    (- +) & (- +) & (- +) & (- -) & (- -) & (+ +) \\ \hline
    (+ -) & (+ -) & (+ -) & (+ +) & (+ +) & (- -)
  \end{array} \right).
\label{eq:phi6trans}
\end{equation}
Hence,  we have obtained doublet-triplet splitting via orbifolding.

\subsection{$D_4$ Family Symmetry}

\begin{figure*}
 \scalebox{0.85} {
  \includegraphics{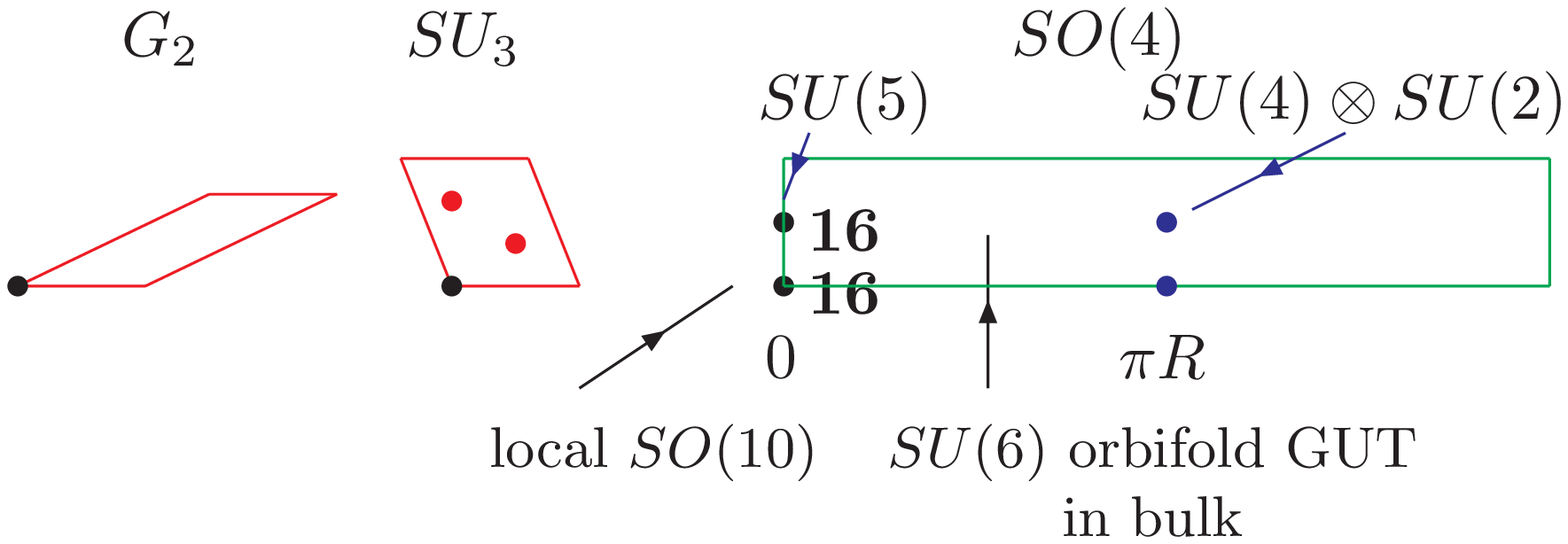}}
  \caption{The two families in the $T_1$ twisted sector.   \label{fig:2fam}}
\end{figure*}

Consider the $\mz_2$ fixed points.   We have four fixed points,
separated into an $SU(5)$ and SM invariant pair by the $W_2$ Wilson
line (see Fig. \ref{fig:2fam}).   We find two complete families, one
on each of the $SO_{10}$ fixed points and a small set of vector-like
exotics (with fractional electric charge) on the other fixed points.
Since $W_2$ is in the direction orthogonal to the two families, we
find a non-trivial $D_4$ family symmetry. This will affect a
possible hierarchy of fermion masses. We will discuss the family
symmetry and the exotics in more detail next.

The discrete group $D_4$ is a non-abelian discrete subgroup of
$SU_2$ of order 8.  It is generated by the set of $2 \times 2$ Pauli
matrices
\begin{equation}
D_4 =  \{ \pm 1, \pm \sigma_1, \pm \sigma_3,  \mp i \sigma_2 \}.
\end{equation}
In our case, the action of the transformation  $\sigma_1 = \left(
\begin{array}{cc} 0 & 1 \\ 1 & 0 \end{array} \right)$ takes $F_1
\leftrightarrow F_2$,  while the action of $\sigma_3 = \left(
\begin{array}{cc} 1 &0 \\ 0 & -1
\end{array} \right)$ takes  $F_2 \rightarrow - F_2$.   These are symmetries of the string.  The first is an
unbroken part of the translation group in the direction orthogonal
to $W_2$ in the $SO_4$ torus and the latter is a stringy selection
rule resulting from $\mz_2$ space group invariance.   Under $D_4$
the three families of quarks and leptons transform as a doublet,
($F_1, \ F_2$), and a singlet, $F_3$.  Only the third family can
have a tree level Yukawa coupling to the Higgs (which is also a
$D_4$ singlet).
In summary:
\begin{itemize}
\item Since the top quarks and the Higgs are derived from the $SU(6)$ chiral adjoint and {\bf 20} hypermultiplet in the 5D
bulk, they have a tree level Yukawa interaction given by
\begin{equation}
W \supset \frac{g_5}{\sqrt{\pi R}} \int_0^{\pi R} dy {\bf {20^c}} \
\Sigma \ {\bf 20} = g_G \ Q_3 \ H_u \ U^c_3
\end{equation}
where $g_5$ ($g_G$) is the 5d (4d) $SU(6)$ gauge coupling constant
evaluated at the string scale.

\item The first two families reside at the $\mz_2$ fixed points, resulting in a $D_4$ family symmetry.
Hence family symmetry breaking may be used to generate a hierarchy
of fermion masses.\footnote{For a discussion of $D_4$ family
symmetry and phenomenology, see Ref. \cite{Ko:2007dz}.  For a
general discussion of discrete non-Abelian family symmetries from
orbifold compactifications of the heterotic string, see
\cite{Kobayashi:2006wq}.}

\end{itemize}

\subsection{More details of ``Benchmark" Model 1 \cite{Lebedev:2007hv}}

Let us now consider the spectrum, exotics, R parity, Yukawa
couplings, and neutrino masses.   In Table \ref{tab:states} we list
the states of the model.   In addition to the three families of
quarks and leptons and one pair of Higgs doublets, we have
vector-like exotics (states which can obtain mass without breaking
any SM symmetry) and SM singlets.   The SM singlets enter the
superpotential in several important ways.   They can give mass to
the vector-like exotics via effective mass terms of the form \be E
E^c \tilde S^n  \ee  where  $E, E^c$ ($\tilde S$) represent the
vector-like exotics and SM singlets respectively.   We have checked
that all vector-like exotics obtain mass at supersymmetric points in
moduli space with $F = D = 0$.   The SM singlets also generate
effective Yukawa matrices for quarks and leptons, including
neutrinos. In addition, the SM singlets give Majorana mass to the 16
right-handed neutrinos $n^c_i$, 13 conjugate neutrinos $n_i$ and
Dirac mass mixing the two. We have checked that the theory has only
3 light left-handed neutrinos.

However, one of the most important constraints in this construction
is the existence of an exact low energy R parity. In this model we
identified a generalized \BmL (see Table \ref{tab:states}) which is
standard for the SM states and vector-like on the vector-like
exotics.  This $\BmL$ naturally distinguishes the Higgs and lepton
doublets. Moreover we found SM singlet states \be \tilde S = \{ h_i,
\ \chi_i, \ s_i^0 \} \ee which can get vacuum expectation values
preserving a matter parity $\mathbb{Z}^{\cal M}_2$ subgroup of
$U(1)_{\BmL}$.  It is this set of SM singlets which give vector-like
exotics mass and effective Yukawa matrices for quarks and leptons.
In addition, the states $\chi_i$ give Majorana mass to neutrinos.

As a final note,  we have evaluated the $\mu$ term in this model. As
a consequence of gauge-Higgs unification, the product  $H_u  H_d$ is
a singlet under all $U(1)$s.   Moreover, it is also invariant under
all string selection rules, i.e.  H-momentum and space-group
selection constraints.   As a result the $\mu$ term is of the form
\begin{equation}   \mu \ H_u \ H_d = W_0(\tilde S) \ H_u \ H_d  \end{equation} where
the factor $W_0(\tilde S)( = \mu)$ is a polynomial in SM singlets
and includes all terms which can also appear in the superpotential
for the SM singlet fields, $\widetilde{W}_0(\tilde S)$. Thus when we
demand a flat space supersymmetric limit, we are also forced to $\mu
\equiv W_0(\tilde S) = \widetilde{W}_0{\tilde S} = 0$\footnote{We
have not shown that the coefficients of the individual monomials in
$W_0(\tilde S)$ are, in general, identical in both the $\mu$ term
and in the SM singlet superpotential term, $\widetilde{W}_0{\tilde
S}$. Nevertheless at 6th order in SM singlet fields we have shown
that when one vanishes, so does the other. This is because each
monomial contains a bi-linear in $D_4$ doublets and this family
symmetry fixes the relative coefficient in the product. Therefore
when the product of $D_4$ doublets vanishes, we have $\mu =
W_0(\tilde S) = \widetilde{W}_0{\tilde S} = 0$}, i.e. $\mu$ vanishes
in the flat space supersymmetric limit. This is encouraging, since
when SUSY is broken we expect both terms to be non-vanishing and of
order the weak scale.

\begin{table}[h]
\centerline{
\begin{tabular}{|c|l|l|c|c|l|l|}
\hline
\# & irrep & label & & \# & irrep & label\\
\hline
 3 &
$\left(\boldsymbol{3},\boldsymbol{2};\boldsymbol{1},\boldsymbol{1}\right)_{(1/3,1/3)}$
 & $q_i$
 & &
 3 &
$\left(\overline{\boldsymbol{3}},\boldsymbol{1};\boldsymbol{1},\boldsymbol{1}\right)_{(-4/3,-1/3)}$
 & $u^c_i$
 \\
 3 &
$\left(\boldsymbol{1},\boldsymbol{1};\boldsymbol{1},\boldsymbol{1}\right)_{(2,1)}$
 & $e^c_i$
 & &
 8 &
$\left(\boldsymbol{1},\boldsymbol{2};\boldsymbol{1},\boldsymbol{1}\right)_{(0,*)}$
 & $m_i$
 \\
 4 &
$\left(\overline{\boldsymbol{3}},\boldsymbol{1};\boldsymbol{1},\boldsymbol{1}\right)_{(2/3,-1/3)}$
 & $d^c_i$
 & &
 1 &
$\left(\boldsymbol{3},\boldsymbol{1};\boldsymbol{1},\boldsymbol{1}\right)_{(-2/3,1/3)}$
 & $d_i$
 \\
 4 &
$\left(\boldsymbol{1},\boldsymbol{2};\boldsymbol{1},\boldsymbol{1}\right)_{(-1,-1)}$
 & $\ell_i$
 & &
 1 &
$\left(\boldsymbol{1},\boldsymbol{2};\boldsymbol{1},\boldsymbol{1}\right)_{(1,1)}$
 & $\ell^c_i$
 \\
 1 &
$\left(\boldsymbol{1},\boldsymbol{2};\boldsymbol{1},\boldsymbol{1}\right)_{(-1,0)}$
 & $\phi_i$
 & &
 1 &
$\left(\boldsymbol{1},\boldsymbol{2};\boldsymbol{1},\boldsymbol{1}\right)_{(1,0)}$
 & $\phi^c_i$
 \\
 6 &
$\left(\overline{\boldsymbol{3}},\boldsymbol{1};\boldsymbol{1},\boldsymbol{1}\right)_{(2/3,2/3)}$
 & $\delta^c_i$
 & &
 6 &
$\left(\boldsymbol{3},\boldsymbol{1};\boldsymbol{1},\boldsymbol{1}\right)_{(-2/3,-2/3)}$
 & $\delta_i$
 \\
 14 &
$\left(\boldsymbol{1},\boldsymbol{1};\boldsymbol{1},\boldsymbol{1}\right)_{(1,*)}$
 & $s^+_i$
 & &
 14 &
$\left(\boldsymbol{1},\boldsymbol{1};\boldsymbol{1},\boldsymbol{1}\right)_{(-1,*)}$
 & $s^-_i$
 \\
 16 &
$\left(\boldsymbol{1},\boldsymbol{1};\boldsymbol{1},\boldsymbol{1}\right)_{(0,1)}$
 & $n^c_i$
 & &
 13 &
$\left(\boldsymbol{1},\boldsymbol{1};\boldsymbol{1},\boldsymbol{1}\right)_{(0,-1)}$
 & $n_i$
 \\
 5 &
$\left(\boldsymbol{1},\boldsymbol{1};\boldsymbol{1},\boldsymbol{2}\right)_{(0,1)}$
 & $\eta^c_i$
 & &
 5 &
$\left(\boldsymbol{1},\boldsymbol{1};\boldsymbol{1},\boldsymbol{2}\right)_{(0,-1)}$
 & $\eta_i$
 \\
 10 &
$\left(\boldsymbol{1},\boldsymbol{1};\boldsymbol{1},\boldsymbol{2}\right)_{(0,0)}$
 & $h_i$
 & &
 2 &
$\left(\boldsymbol{1},\boldsymbol{2};\boldsymbol{1},\boldsymbol{2}\right)_{(0,0)}$
 & $y_i$
 \\
 6 &
$\left(\boldsymbol{1},\boldsymbol{1};\boldsymbol{4},\boldsymbol{1}\right)_{(0,*)}$
 & $f_i$
 & &
 6 &
$\left(\boldsymbol{1},\boldsymbol{1};\overline{\boldsymbol{4}},\boldsymbol{1}\right)_{(0,*)}$
 & $f^c_i$
 \\
 2 &
$\left(\boldsymbol{1},\boldsymbol{1};\boldsymbol{4},\boldsymbol{1}\right)_{(-1,-1)}$
 & $f_i^-$
 & &
 2 &
$\left(\boldsymbol{1},\boldsymbol{1};\overline{\boldsymbol{4}},\boldsymbol{1}\right)_{(1,1)}$
 & ${f^c_i}^+$
 \\
 4 &
$\left(\boldsymbol{1},\boldsymbol{1};\boldsymbol{1},\boldsymbol{1}\right)_{(0,\pm2)}$
 & $\chi_i$
 & &
 32 &
$\left(\boldsymbol{1},\boldsymbol{1};\boldsymbol{1},\boldsymbol{1}\right)_{(0,0)}$
 & $s^0_i$
 \\
 2 &
$\left(\overline{\boldsymbol{3}},\boldsymbol{1};\boldsymbol{1},\boldsymbol{1}\right)_{(-1/3,2/3)}$
 & $v^c_i$
 & &
 2 &
$\left(\boldsymbol{3},\boldsymbol{1};\boldsymbol{1},\boldsymbol{1}\right)_{(1/3,-2/3)}$
 & $v_i$
 \\
\hline
\end{tabular}
} \caption{Spectrum. The quantum numbers under
$\SU3\times\SU2\times[\SU4\times\SU2']$ are shown in boldface;
hypercharge and \BmL\ charge appear as subscripts.  Note that the
states $s_i^\pm$, $f_i$, $\bar f_i$ and $m_i$ have different $B-L$
charges for different $i$, which we do not explicitly list
\cite{Lebedev:2007hv}.} \label{tab:states}
\end{table}

\subsection{Gauge Coupling Unification and Proton Decay}

\begin{figure*}[ht!]
         \scalebox{0.55} {
        \includegraphics{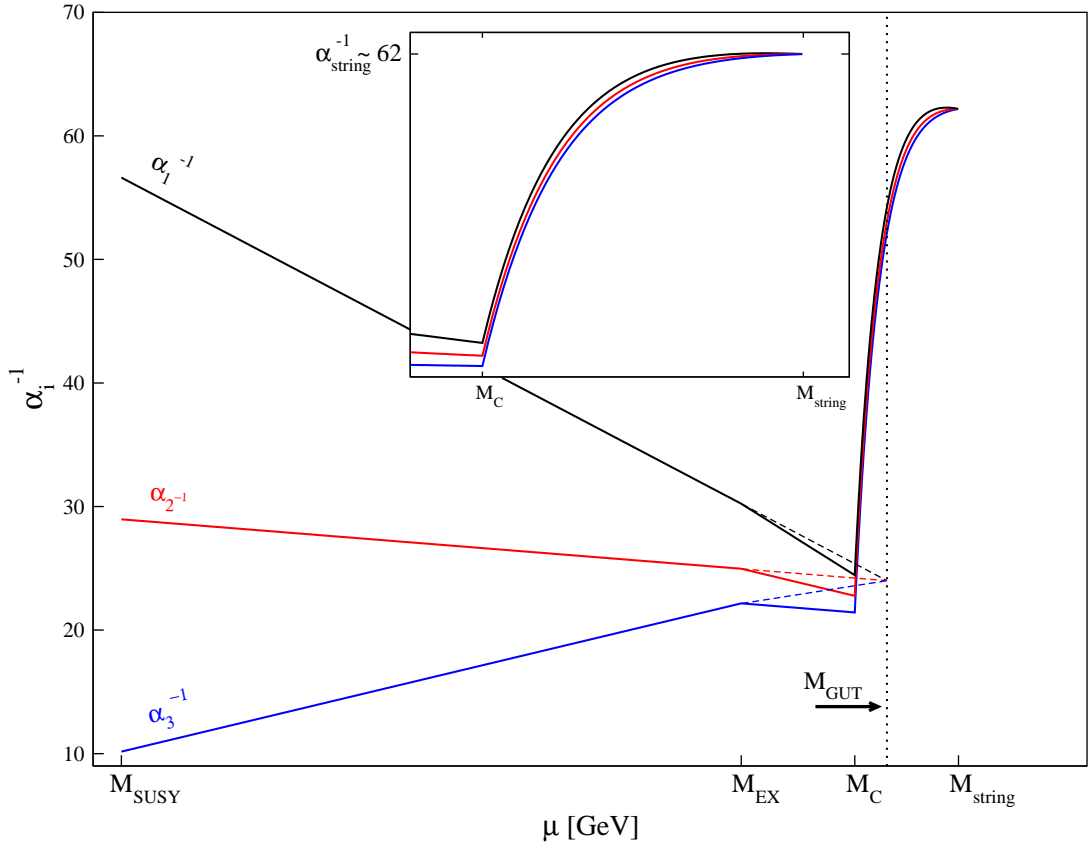}}
        \caption{An example of the type of gauge coupling evolution we see in these models,
        versus the typical behavior in the MSSM.  The ``tail'' is due to the power law running
        of the couplings when towers of Kaluza-Klein modes are involved.  Unification in this
        model occurs at $M_{\rm STRING} \simeq 5.5\times 10^{17}$ GeV, with a compactification scale
        of $M_{c} \simeq 8.2 \times 10^{15}$ GeV, and an exotic mass scale of $M_{\rm EX} \simeq 8.2 \times 10^{13}$ GeV.}
        \label{fig:beta_functions}
\end{figure*}

We have checked whether the SM gauge couplings unify at the string
scale in the class of models similar to Model 1 above
\cite{Dundee:2008ts}.  All of the 15 MSSM-like models of Ref.
\cite{Lebedev:2007hv} have 3 families of quarks and leptons and one
or more pairs of Higgs doublets. They all admit an $SU(6)$ orbifold
GUT with gauge-Higgs unification and the third family in the bulk.
They differ, however, in other bulk and brane exotic states.   We
show that the KK modes of the model, including only those of the
third family and the gauge sector, are {\em not} consistent with
gauge coupling unification at the string scale.  Nevertheless, we
show that it is possible to obtain unification if one adjusts the
spectrum of vector-like exotics below the compactification scale. As
an example, see Fig. \ref{fig:beta_functions}.  Note, the
compactification scale is less than the 4d GUT scale and some
exotics have mass two orders of magnitude less than $M_{c}$, while
all others are taken to have mass at $M_{\rm STRING}$.   In
addition,  the value of the GUT coupling at the string scale,
$\alpha_G(M_{STRING}) \equiv \alpha_{string}$, satisfies the weakly
coupled heterotic string relation \be \label{het_string_BC} G_N =
\frac{1}{8}\,\alpha_{string}\,\alpha^\prime \ee  or  \be
\alpha_{string}^{-1} = \frac{1}{8}\, (\frac{M_{Pl}}{M_{\rm
STRING}})^2 . \ee

%~~~~~~~~~~~~~~~~~~~~~~~~~~~~~~~~~~~~~~~~~~~~~~~~~~~~~~~   FIGURE beta functions
%~~~~~~~~~~~~~~~~~~~~~~~~~~~~~~~~~~~~~~~~~~~~~~~~~~~~~~~   FIGURE beta functions

\begin{figure*}[ht!]
         \scalebox{0.55} {
        \includegraphics{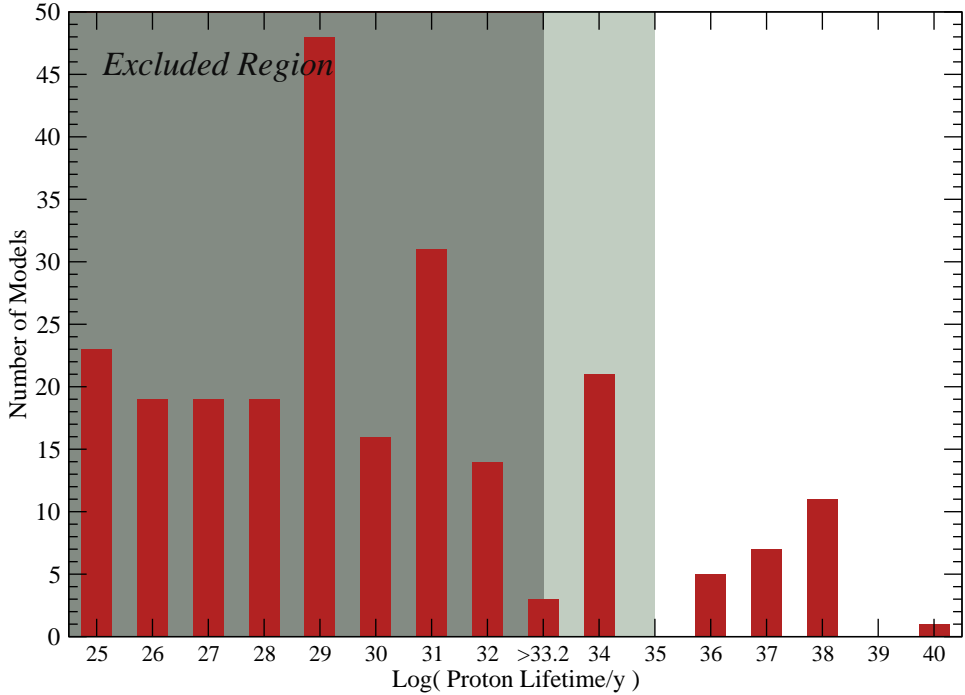}}
        \caption{Histogram of solutions with $M_{\rm STRING} > M_{c} \gtrsim M_{\rm EX}$, showing the
        models which are excluded by Super-K bounds (darker green) and those which are
        potentially accessible in a next generation proton decay experiment (lighter green).
        Of 252 total solutions, 48 are not experimentally ruled out by the current
        experimental bound, and most of the remaining parameter space can be eliminated
        in the next generation of proposed proton decay searches.}
        \label{fig:histogram}
\end{figure*}

In Fig. \ref{fig:histogram} we plot the distribution of solutions
with different choices of light exotics.   On the same plot we give
the proton lifetime due to dimension 6 operators.  Recall in these
models the two light families are located on the $SU(5)$ branes,
thus the proton decay rate is only suppressed by $M_{c}^{-2}$. Note,
90\% of the models are already excluded by the Super-Kamiokande
bounds on the proton lifetime.  The remaining models may be tested
at a next generation megaton water \v{c}erenkov detector.

%~~~~~~~~~~~~~~~~~~~~~~~~~~~~~~~~~~~~~~~~~~~~~~~~~~~ FIGURE histogram
%~~~~~~~~~~~~~~~~~~~~~~~~~~~~~~~~~~~~~~~~~~~~~~~~~~~~~~~   FIGURE histogram

\section{Conclusion}

In these lectures we have discussed an evolution of SUSY GUT model
building.   We saw that 4d SUSY GUTs have many virtues.  However
there are some problems which suggest that these models may be
difficult to derive from a more fundamental theory, i.e. string
theory.   We then discussed orbifold GUT field theories which solve
two of the most difficult problems of 4d GUTs, i.e. GUT symmetry
breaking and Higgs doublet-triplet splitting.   We then showed how
some orbifold GUTs can find an ultra-violet completion within the
context of heterotic string theory.

The flood gates are now wide open.  In recent work
\cite{Lebedev:2007hv} we have obtained many models with features
like the MSSM:  SM gauge group with 3 families and vector-like
exotics which can, in principle, obtain large mass.   The models
have an exact R-parity and non-trivial Yukawa matrices for quarks
and leptons.  In addition, neutrinos obtain mass via the See-Saw
mechanism.   We showed that gauge coupling unification can be
accommodated \cite{Dundee:2008ts}.   Recently, another MSSM-like
model has been obtained with the heterotic string compactified on a
$T^6/\mz_{12}$ orbifold \cite{Kim:2007mt}.

Of course, this is not the end of the story.  It is just the
beginning.   We must still obtain predictions for the LHC.  This
requires stabilizing the moduli and breaking supersymmetry.  In
fact, these two conditions are {\em not} independent, since once
SUSY is broken, the moduli will be stabilized.  The scary fact is
that the moduli have to be stabilized at just the right values to be
consistent with low energy phenomenology.

\section*{Acknowledgments}
This work is supported under DOE grant number DOE/ER/01545-879.

\end{document}